\documentclass[reprint,aps,prd,twocolumn,showpacs,showkeys, 10pt, longbibliography, nolinenumbers, floatfix]{revtex4-1}

\usepackage{amsmath}
\usepackage{amssymb}
\usepackage{amsfonts}
\usepackage{graphicx}
\usepackage{float}
\usepackage{dcolumn}
\usepackage{multirow}
\usepackage{natbib}
\usepackage{tcolorbox}
\usepackage{parskip}
\usepackage[colorlinks = true,linkcolor = blue,urlcolor  = blue,citecolor = blue,anchorcolor = blue]{hyperref}
\usepackage{enumerate}

\usepackage{times}

\begin{document}


\title{Hybrid Stars with a Crossover Equation of State}
\author{G.A. Tsalis and Ch.C. Moustakidis}



\affiliation{Department of Theoretical Physics, Aristotle University of Thessaloniki, 54124 Thessaloniki, Greece 
}
\begin{abstract}
The possible occurrence of a phase transition in superdense nuclear matter remains an open problem in nuclear physics. In particular, the transition from purely hadronic matter to the hypothesized deconfined quark matter is a subject of considerable interest. Beyond its fundamental significance, this transition has important implications for the physics of compact objects, including neutron stars, quark stars, and hybrid stars. The conventional approaches to this problem involve the Maxwell and Gibbs constructions, both of which have been extensively studied and applied. In the present study, we focus on a third, alternative approach based on a smooth transition from hadronic to quark matter through the so-called switching function. More specifically, the present work focuses on two main objectives. The first is to systematically investigate the influence of the symmetry energy on the transition densities and chemical potentials associated with the transition to quark matter. Subsequently, we attempt to physically constrain and parametrize the two parameters characterizing the switching function, ensuring that the transition from hadronic to quark matter is not merely continuous but also consistent with relevant physical requirements. The main conclusion is that, with an appropriate choice of the aforementioned parameters, the phenomenological model becomes more physically motivated and can serve as a useful tool for investigating the structure and properties of hybrid compact objects. Moreover, this approach offers the flexibility to predict the properties of compact objects that lie beyond the scope of standard neutron-star theory, such as objects characterized by relatively low masses and small radii, which have recently attracted considerable attention in observational studies.


\end{abstract}

\maketitle


\section{Introduction}
One of the fundamental challenges in Nuclear Physics and Astrophysics is to understand the composition of dense nuclear matter and characterize its essential properties at both zero and finite temperatures~\cite{Glendenning1996CompactSN,Shapiro:1983du,Haensel2007NeutronS1,schaffner-bielich_2020}.  The difficulty in understanding the properties of superdense nuclear matter stems not only from the lack, or at least the limited availability, of experimental data, but also from the absence of a sound and robust theoretical framework capable of adequately accounting for the composition of this superdense matter and its corresponding degrees of freedom. In particular, the problem of the phase transition from hadronic matter to deconfined quark matter, which represents an energetically allowed process, is of considerable importance and should be carefully considered in any systematic investigation of dense nuclear matter~\cite{Fukushima-2011,Baym-2018,Glendenning-1992,Baym-2019,Schafer-1999,Schafer-1999(2),Schertler-2000,Glendenning-2001,Ju-2021,Curin-2021,Masuda-2013,Masuda-2016,Patra-2026,Lopes-2024,Zhao-2022,Ju-2025,Abgaryan-2018,Fraga-2019,Lugones-2013,Schmitt-2020,Palhares-2010,Wang-2022,Blacker-2023,Espino-2022,Naseri-2024,Laskos-2025,Laskos-2024,Tsaloukidis-2023,Guo-2025}. 

A possible phase transition in the interior of neutron stars can have profound consequences for their structure and fundamental properties. In particular, the emergence of such a hybrid state of matter modifies the equation of state (EoS) of dense nuclear matter and, consequently, the key macroscopic properties of neutron stars, including their masses, radii, tidal deformabilities, and related observables.
Furthermore, phase transitions are expected to affect the characteristic frequencies associated with various modes of stellar oscillation, including r-modes, f-modes, p-modes, and g-modes and so on. They may also influence the thermal evolution and cooling of neutron stars, while potentially playing a crucial role in phenomena associated with their rapid rotation and temporal evolution, such as pulsar glitches.
The concept of twin stars is closely related to the occurrence of such phase transitions. Consequently, the potential identification of twin-star configurations would provide strong evidence for the existence of phase transitions in the dense-matter interior of neutron stars~\cite{Alford2005}.

These phase transitions have been extensively studied within conventional theoretical frameworks, most notably through the Maxwell and Gibbs constructions, which have been widely investigated and applied in the literature~\cite{Baym-2018,Constantinou-2023}.  However, a relatively recent alternative method has been proposed which, in a sense, combines the aforementioned approaches and provides a phenomenological treatment of the problem~\cite{KapustaWelle2021,Kapusta-Welle2022,Kapusta2010,Albright-2015,AlbrightKapusta2014}.  In particular, they  begin by formulating the pressure $P$ in terms of a hadronic contribution $P_h$, a perturbative QCD contribution $P_{\mathrm{qg}}$, and a phenomenological switching function $S(T,\mu)$, which interpolates between the two regimes. The hadronic contribution $P_h$ may be evaluated within any of the established hadronic models available in the literature as well as the quark-gluon pressure  $P_{\mathrm{qg}}$ can be obtained from an appropriate quark–gluon model. The switching function $S(T,\mu)$ is required to vanish in the low-temperature and low-chemical-potential regime and approach unity in the high-temperature and high-chemical-potential regime. In addition, $S(T,\mu)$ must exhibit sufficient smoothness to ensure that no artificial first-, second-, or higher-order phase transitions are introduced~\cite{KapustaWelle2021,Kapusta-Welle2022,Kapusta2010,Albright-2015,AlbrightKapusta2014}. The function $S(\mu,T)$ acts as a switching function that varies continuously between 0 and 1 over appropriate ranges of the chemical potential $\mu$ and temperature $T$, thereby ensuring a smooth transition from hadronic matter to deconfined quark matter while simultaneously accounting for thermal effects. It is required to be an even function of $\mu$ and infinitely differentiable, thereby ensuring that no artificial phase transition of any order is introduced. 

The aforementioned approach has been applied to the study of neutron stars, based on the assumption that the transition from cold nuclear or neutron matter to quark matter may likewise occur through a crossover~\cite{Kapusta-Welle2022}.  It has also been used to investigate its general implications of such a transition for the structure and properties of neutron stars~\cite{BlaschkeHanu-2022,Ayriyan-2021}. Having developed such a method, one can investigate hybrid equations of state in a relatively simple yet highly effective manner, as well as systematically explore their consequences for the corresponding properties of hybrid stars. 

The aim of the present work is twofold. First, we perform a systematic investigation of the impact of the nuclear symmetry energy on the densities at which the phase transition from hadronic matter to deconfined quark matter takes place. To this end, the hadronic equation of state is appropriately parametrized, while an extended set of CFL equations of state~\cite{Alford2005,flores2017constraining,lugones2002color,lugones2003high,horvath2004self,Alford-2001,oikonomou2023color} is employed to describe the deconfined quark matter. Particular emphasis is placed not only on determining the critical baryon density at which the phase transition occurs, but also on identifying the corresponding chemical potential at the transition. Having calibrated the resulting equations of state in this manner, they can now be regarded as physics-informed and are subsequently employed to investigate fundamental properties of neutron stars, including hybrid stars, such as their masses, radii, and tidal deformabilities, and to establish correlations between these properties and the chosen parameterization. Particular emphasis is also placed on the study of the speed of sound, which serves as a key indicator of the behavior of nuclear matter at both low and high densities. The possibility of applying the present approach to explain recently observed exotic compact objects characterized by relatively low masses and small radii, which lie beyond the predictions of the standard theory of neutron stars, is discussed and analyzed.

The paper is organized as follows. Section II presents the equation of state of hadronic matter, while Section III is devoted to the formulation of the equation of state for deconfined quark matter. In Section IV, we introduce and discuss the crossover equation of state. Section V briefly reviews the Tolman–Oppenheimer–Volkoff (TOV) equations and the tidal deformability. The main results are presented and discussed in Section VI. Finally, Section VII summarizes the principal findings of the study and concludes with some final remarks.


\section{Equation of state of  hadronic matter}

For the description of hadronic matter, we will rely on a relatively simple yet robust model. In particular, we construct the hadronic EoS by employing an expansion of the energy per nucleon as a function of the baryon density $n$ around the nuclear saturation density $n_0$, given by~\cite{DivarisMoustachidis-2024,Lattimer-2023}
\begin{equation} \label{1}
    E(n,\alpha) = E_0 + \frac{K_0}{18n_0^2}(n-n_0)^2 +S(n)\alpha^2,
\end{equation}
where $\alpha=\frac{n_n-n_p}{n}$ denotes the asymmetry parameter, while $n_n$ and $n_p$ represent the number densities of neutron and protons, respectively. The parameters $E_0$ and $K_0$ denote the energy per baryon of symmetric nuclear matter and the incompressibility at $n_0$, respectively. Finally, $S(n)$ is the well-known symmetry energy~\cite{Lattimer-2023,Baldo-2016,Steiner-2005,Sammarruca-2013,Bao-2008,Horowitz-2014}, which can also be expanded around saturation density $n_0$ leading to ~\cite{DivarisMoustachidis-2024,Lattimer-2023,Piekarewicz-2009,Lattimer-2014}
\begin{equation} 
\label{2}
   S(n) = J + \frac{L}{3n_0}(n-n_0)+\frac{K_{sym}}{18n_0^2}(n-n_0)^2 + ...,
\end{equation}
where $J$ denotes the symmetry energy at saturation density $J = S(n_0)$ and $L$ denotes the so-called symmetry energy slope. The parameter $K_{sym}$ represents the curvature of symmetry energy, which is not well constraint, while $L$ represents the slope parameter. Both parameters are defined by the following derivatives:

\begin{equation}\label{3}
    L = 3n_0\left(\frac{dS(n)}{dn}\right)_{n=n_0}
\end{equation}
\begin{equation}\label{4}
    K_{sym} = 9n_0^2\left(\frac{dS^2(n)}{d^2n}\right)_{n=n_0}
\end{equation}

The derivation of the equation of state relies on enforcing chemical equilibrium and charge neutrality in the system. These conditions read 
\begin{equation} \label{5}
    \mu_n=\mu_p+\mu_e \hspace{0.5 cm}\mathrm{and} \hspace{0.5 cm} n_p=n_e,
\end{equation}
where $\mu_i$ denotes the chemical potential of the different particle species and $n_e$ is the electron number density. Solving Equation~(\ref{5}) yields the parameters $\alpha$ and  $n_e$ as functions of the baryon density. The resulting equation of state is then obtained by simply summing the contributions of baryons and electrons to the energy density and pressure. The outermost layers of the stellar models are described by the well-known Feynman-Metropolis ans Teller EoS~\cite{Feynman-1949}. The transition density from the uniform core to the the solid crust is calculated via the Thermodynamical method~\cite{Moustakidis-2010}.

The application of the aforementioned nuclear model of hadronic matter in hybrid stars, i.e., the transition from finite nuclei to nuclear matter of hybrid stars, is achieved through the quantity $\varepsilon_b = nE$. The resulting expression for the energy density as function of baryon density reads:
\begin{equation} \label{6}
    \begin{aligned}
        \varepsilon_b(n,\alpha) = E_0n\ +\ \frac{K_0}{18n_0^2}n\left(n-n_0\right)^2 \\ 
        +\ \left(J\ +\ \frac{L}{3n_0}\left(n-n_0\right) \right)n\alpha^2
    \end{aligned}
\end{equation} 

According to Refs.~\cite{DivarisMoustachidis-2024,Kanakis-2026}, there are two main reasons for extending this hadronic model to star matter. First, it can be applied to neutron-star matter because the second-order expansion in the asymmetry parameter $\alpha$ is also applicable to neutron-rich nuclei, which are present in neutron star-matter. Second, the model depends directly on the two parameters $K_0,\ L$, or, equivalently, on the parameter $\eta\, =\, \left(K_0L^2\right)^{1/3}$ (for more details see Refs.~\cite{Sotani-2014,Sotani-2022}).  Consequently, a variety of equations of state, ranging from softer to stiffer ones, can be constructed. 

\begin{table}[h]
\caption{The incompressibility $K_0$ (in MeV), the slope parameter $L$ (in MeV) and the corresponding values of  the parameter $\eta$ (in MeV) used in the present study. }
\begin{tabular}{ c c c cccccccc }
\hline
$ K_0 $	& 220& 224 &228& 232&  236& 240 & 244& 248&  252&  256  \\
\hline
$L$ & 40 &48 & 56&  64&  72& 80& 88&  96&  104&  112   \\
\hline
$\eta$ & 70.6 &80.2& 89.4& 98.3& 107.0&115.4& 123.6& 131.7& 139.7& 147.5 \\
\hline
\end{tabular}
\label{Table-1}
\end{table}

At this point, it should be noted that, in order to construct the hybrid - crossover model, following Refs.~\cite{DivarisMoustachidis-2024,Kanakis-2026} , where the symmetry energy $J$ is fixed at $32\,{\rm MeV}$, while the symmetry energy slope $L$ ranges from $40\,{\rm MeV}$ to $112\, {\rm MeV}$. The  symmetry energy curvature at saturation density, denoted by $K_0$, is varied from 220 to 256 MeV. 
For the parameters $L$, $K_0$ and  $\eta$ we adopt the parametrization used in a previous work~\cite{Divaris-2026} (see also Table~\ref{Table-1}). 

In this way, it becomes possible to study the contribution of these specific and well-established EoS \cite{DivarisMoustachidis-2024}, to the crossover model. Additionally, the baryon density at saturation density is fixed at: $n_0\,=\,0.16\ {\rm fm}^{-3}$, while the energy per baryon at the saturation point is also held constant at $E_0 \, = \, -16\, {\rm MeV}$.

It is noteworthy that the construction of the hadronic model  allows the model to be controlled and modified through a single parameter, namely $\eta$~\cite{DivarisMoustachidis-2024}. In other words, choosing different combinations of $K_0$ and $L$ results in different hadronic equations of state (EoSs), which can be either stiffer or softer, depending on the requirements of the hybrid model.

The pressure, due to the baryons which is  defined as 
\begin{equation} 
P_b=n^2\frac{d({\varepsilon_b}/n)}{dn}
 \label{Pres-def}   
\end{equation}
reads now 
\begin{equation}
P_b=\frac{K_0}{9 n_0^2}n^2  \left(n-n_0\right)+\alpha^2 \frac{L}{3n_0}n^2
\label{Pres-1}    
\end{equation}

The electron contribution to the total energy density and pressure is given by the well-known expressions for a relativistic Fermi gas, namely 
\begin{equation}
{\cal E}_e=\frac{(m_ec^2)^4}{8\pi^2 (\hbar c)^3} \left[z(2z^2+1)\sqrt{1+z^2}-\ln\left( z+\sqrt{z^2+1}\right)  \right]   
\label{Ed-ele}
\end{equation}
and 
\begin{equation}
P_e=\frac{(m_ec^2)^4}{24\pi^2 (\hbar c)^3} \left[z(2z^2-3)\sqrt{1+z^2}+3\ln\left( z+\sqrt{z^2+1}\right)   \right]   
\label{Pre-ele}
\end{equation}
\begin{equation}
z=\frac{(\hbar c) (3\pi^2 n_e)^{1/3}}{m_e c^2}, \quad n_e=x_p n  
\label{xelectron}
\end{equation}
where $x_p$ is the proton fraction and $n_e$ the electron number density. 
The total energy density and pressure of charge-neutral, chemically equilibrated matter are given by
\begin{equation}
 {\cal E}_{\rm h}={\varepsilon}_b+{\cal E}_e
 \label{en-total}
\end{equation}
\begin{equation}
 P_{\rm h}=P_b+P_e
 \label{Pres-total}
\end{equation}
From Eqs.~(\ref{en-total}) and (\ref{Pres-total}) we construct the equation of state for neutron star matter. 
The proton fraction $x_p=n_p/n$, which plays a crucial role in determining neutron star properties, is highly sensitive to the nuclear symmetry energy. In particular, the condition of beta equilibrium in the interior of neutron stars
\begin{equation}
\mu_n=\mu_p+\mu_e   
\label{beta-equil}
\end{equation}
leads to the following equation
\begin{equation}
4(1-2x_p)S(n)=\hbar c (3\pi^2 n_e)^{1/3}=\hbar c (3\pi^2 n x_p)^{1/3}
 \label{cond-1}
\end{equation}
Solving Eq.~(\ref{cond-1}) we found the density dependence of the proton fraction
\begin{equation}
x_p(n)=\frac{1}{2}-\frac{1}{4}\left( [2\beta(\gamma+1)]^{1/3}- [2\beta(\gamma-1)]^{1/3}\right)
\label{sol-1}
\end{equation}
where
\[\beta=3\pi^2n\left(\frac{\hbar c}{4 S(n)}  \right)^3, \qquad \gamma=\left(1+\frac{2\beta}{27}  \right)^{1/2}  \]
%
These ingredients are used to construct the hadronic component of matter, which is then combined with the CFL model, discussed in the following section, to obtain a crossover equation of state. This approach provides an alternative to the conventional Maxwell  and Gibbs constructions for describing a first-order phase transition~\cite{KapustaWelle2021}.

To construct the crossover model, and since all physical quantities in the CFL model, such as pressure and energy density, are expressed as functions of the chemical potential, the corresponding quantities in the hadronic model must also be expressed as functions of the chemical potential. 

\section{Equation of state of deconfined quark matter}

To describe the deconfined quark matter in the hybrid model, we employ the CFL model as presented in \cite{BlaschkeHanu-2022}. Specifically, this approach is based on the quark pressure $P_q$ expressed as a function of the chemical potential $\mu$, as originally proposed by Alford et. al.~\cite{Alford2005} 

\begin{equation}
P_q(\mu) = \frac{\xi_4a_4}{4\pi^2}\left(\frac{\mu}{3}\right)^4 +\\ \frac{\xi_{2a}\Delta^2-\xi_{2b}m_s}{\pi^2}\left(\frac{\mu}{3}\right)^2 + \mu^4_e - B_{\rm eff}
\end{equation}

The CFL phase is characterized by the parameters $\xi_4 = 3$, $\xi_{2a} = 3$, and $\xi_{2b} = 34$. In addition, the quarks are assumed to be massless, i.e., $m_u = m_d = m_s = 0$, and charge neutrality is imposed through the condition $\mu_e = 0$. The coefficient $a_4$ is assumed to hold constant values. Moreover, $\Delta$ denotes the pairing gap parameter, while $B_{\rm eff}$ represents the effective bag constant. Consequently, the quark pressure in the CFL model is expressed as a function of the chemical potential in the following form

\begin{equation}
    P_q(\mu) = \frac{3}{4\pi^2}a_4\left(\frac{\mu}{3}\right)^4 + \frac{3}{\pi^2}\Delta^2\left(\frac{\mu}{3}\right)^2 - B_{\rm eff}
\end{equation}

The quark number density is also expressed as a function of the chemical potential, as follows

\begin{equation}
    n_q(\mu) = 3\frac{\partial P_q(\mu)}{\partial\mu} \\
    = \frac{3}{\pi^2}a_4\left(\frac{\mu}{3}\right)^3 + \frac{6}{\pi^2}\Delta^2\left(\frac{\mu}{3}\right)
\end{equation}

Consequently, the energy density $\varepsilon_q$ expressed as a function of the chemical potential is given by
\begin{eqnarray}
\varepsilon_q(\mu)&=& -P_q(\mu) + \left(\frac{\mu}{3}\right)n_q(\mu) \nonumber \\
&=& \frac{9}{4\pi^2}a_4\left(\frac{\mu}{3}\right)^4 + \frac{3}{\pi^2}\Delta^2\left(\frac{\mu}{3}\right)^3 + B_{\rm eff}
\label{ED-CFL}
\end{eqnarray}

The speed of sound, expressed in dimensionless form, is calculated by using the following relation
\begin{equation}
    c_s^2(\mu) = \frac{dP_q(\mu)}{d\varepsilon_q(\mu)} = \frac{n_q(\mu)}{\mu dn_q(\mu)/d\mu} = \frac{1 + 2(3\Delta)^2/(a_4\mu^2)}{3 + 2(3\Delta)^2/(a_4\mu^2)}
    \label{CFL-speed}
\end{equation}
In the CFL model, the speed of sound is likewise expressed as a function of the chemical potential and provides an indication of the stiffness of the equation of state. Similarly, in the hadronic model, it is derived from 
Ref.~\cite{DivarisMoustachidis-2024}, the CFL model \cite{BlaschkeHanu-2022} can produce various equations of state for superconducting quark matter. Different variants of quark matter can be obtained by appropriately choosing the model parameters. The three parameters that characterize quark matter and determine its EoS are $a_4, \,  \Delta, \, B_{\rm eff}$. These parameters are conventionally listed in this specific order to characterize the quark matter EoS. To construct a quark matter EoS that crosses the hadronic EoS, the aforementioned parameters are assigned values that ensure that hadronic matter is favored at lower chemical potentials, whereas quark matter becomes favored at higher chemical potentials. In Table~\ref{Table-2}, we present several representative sets of parameters for the CFL model that satisfy the aforementioned conditions.

\begin{table}
\centering
\caption{The values of the parametrers of the CFL model.}
\label{tab:1}
\begin{ruledtabular}
\begin{tabular}{lll}
$a_4$ & $\Delta$ (MeV) & $B_{\rm eff}$ (MeV) \\
\hline
0.6 & $0$ & $180$ \\
0.7 & $0$ & $170$ \\
 0.7 & $0$ & $180$ \\
 0.7 & $0$ & $190$ \\
0.8 & $0$ & $170$ \\
0.7 & $100$ & $190$
\end{tabular}
\end{ruledtabular}
\label{Table-2}
\end{table}

\section{Crossover Equation of State}

\subsubsection{The  switching function}

Following the approach proposed by Albright et al.~\cite{AlbrightKapusta2014} a smooth crossover between the hadronic and quark phases is introduced without a thermodynamic phase transition of either first or second order. Conventionally, the transition is constructed in a different manner. Specifically, the hadron resonance gas description is terminated at temperatures below $200\, {\rm MeV}$, while perturbative QCD becomes dominant at temperatures above $200\,{\rm MeV}$. The intermediate region is then interpolated using a function that connects these two phases. However, the interpolation may introduce discontinuities in derivatives of order $n$, potentially leading to a phase transition of the corresponding order. Such discontinuities are undesirable in a crossover approach. Therefore, the switching function must be carefully chosen to ensure sufficient smoothness and avoid the emergence of first-order, second-order, or higher-order phase transitions~\cite{AlbrightKapusta2014}. 

As shown in \cite{KapustaWelle2021}, various types of switching functions have been developed to achieve a smooth crossover between a hadron gas and deconfined quark matter. However, Albright et al.~\cite{AlbrightKapusta2014} proposed the following switching function, which is employed in the model analyzed in this paper
\begin{equation}
S \left(T,\mu \right)= \exp\left[-
\theta\left(T,\mu \right) \right],
\end{equation}
where the function in the exponent is given by the following expression
\begin{equation}
\theta \left( T\,,\mu \right)\, =\, \left[ \left(\frac{T}{T_0} \right)^q \,+\,\left(\frac{\mu}{\mu_0} \right)^r \right]^{-1}
\end{equation}

This switching function is an exponential function of the temperature $T$ and the chemical potential $\mu$. Exponents $q$ and $r$ are integers, while the parameters $T_0$ and $\mu_0$, govern the crossover between the two phases. The crossover transition becomes narrower at lower temperatures and for higher values of $r$. The main advantage of this switching function is that it is even function and is infinitely differentiable, ensuring that no artificial phase transition of any order is introduced. As a first approximation, Albright et al.~\cite{AlbrightKapusta2014} considered the case $T=0$.  In the present study, we adopt the same approach, assuming a temperature of $T=0$ MeV, while leaving the extension to finite temperatures for future work. To account for color-superconducting effects, which are expected to play an important role at low temperatures, we introduce the diquark pairing gap parameter $\Delta$ and the effective bag parameter $B_{\rm eff}$ into the crossover equation of state. Thus, under this assumption, the switching function reduces to the following form
\begin{equation}
\label{SF-1a}
S \left( \mu \right)= \exp\left[-
\left( \frac{\mu_0}{\mu}\right)^r \right]
\end{equation}

\begin{figure}[ht]
    \centering
    \includegraphics[width=1.0\linewidth]{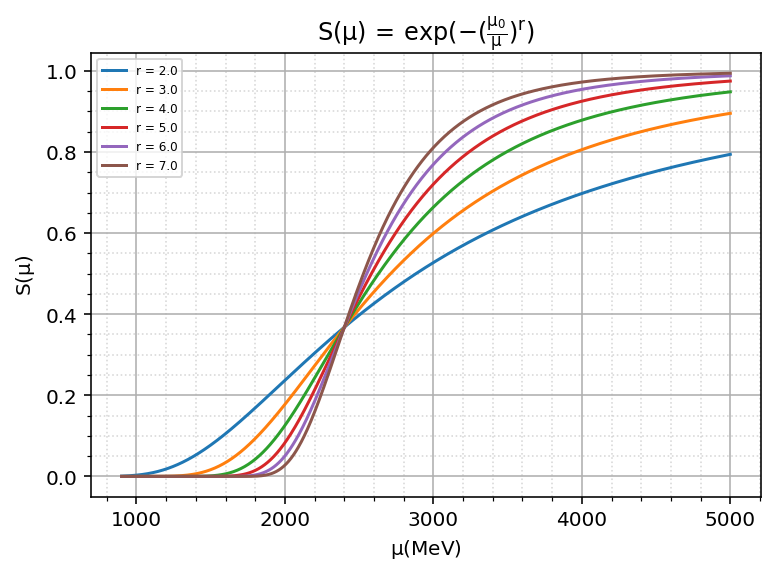}
    \caption{The switching function $S(\mu)$ as a function of the chemical potential $\mu$ for various values of $r$, with $\mu_0=2400\,\rm{MeV}$ held fixed.}
    \label{fig:switchfunction}
\end{figure}

\begin{figure}[ht]
\centering
\includegraphics[width=1.0\linewidth]{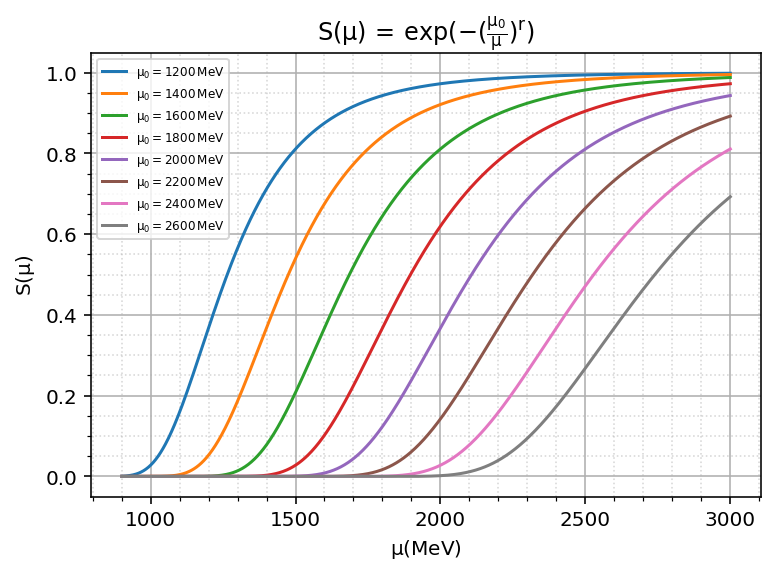}
    \caption{The switching function $S(\mu)$ as a function of the chemical potential $\mu$ for various values of $\mu_0$, with $r=7$ held fixed.}
    \label{fig:switchfunctonr2}
\end{figure}

The switching function is plotted in Fig.~\ref{fig:switchfunction} as a function of the chemical potential $\mu$ for several values of the parameter $r$, while keeping $\mu_0$ fixed. The value of $\mu_0$ determines how rapidly the switching function approaches its asymptotic maximum value. Similarly, the switching function is plotted as a function of $\mu$ for several values of $\mu_0$, while keeping $r$ fixed in Fig. \ref{fig:switchfunctonr2}. In this case, smaller values of $\mu_0$ cause the switching function to approach its asymptotic maximum value more rapidly.

\subsubsection{Crossover equations}
To construct the crossover EoS, the equations governing the hadronic and CFL phases, as formulated in the preceding sections, must be expressed in terms of a common independent variable, namely, the chemical potential $\mu$\, which is defined by the following expression (see also Ref.~\cite{BlaschkeHanu-2022})
\begin{equation}
\mu=\frac{P\,+\,\varepsilon}{n}
\label{chemikal-1}
\end{equation}

The pressure, baryon density, and energy density of the crossover EoS are given by the following equations, respectively:

\begin{equation}\label{PressureCross}
P(\mu)\,=\,S(\mu)P_q(\mu)\,+\,[1-S(\mu)]P_h(\mu)
\end{equation}

\begin{eqnarray}
n(\mu)=\frac{dP(\mu)}{d\mu}&=&S(\mu)n_q(\mu)\,+\,[1-S(\mu)]n_h(\mu)\nonumber\\
&+&\,S^{'}(\mu)[P_q(\mu)\,-\,P_h(\mu)]
\label{BaryonDensity}
\end{eqnarray}
\begin{eqnarray}
E(\mu)&=&-P(\mu)+\mu n(\mu)
\nonumber\\
&=&
S(\mu)\varepsilon_q(\mu)+[1-S(\mu)]{\cal E}_{\rm h} \nonumber\\
&+&r \left( \frac{\mu_0}{\mu} \right)^{r+1}S(\mu)[P_q(\mu)-P_h(\mu)]
\label{EnergyDensityCross}
\end{eqnarray}
where the  switching function $S(\mu)$ is given by Eq. (\ref{SF-1a}), while  its derivative with respect with $\mu$ is given by
\begin{equation}
 S^{'}(\mu)=r\frac{\mu_0^r}{\mu^{r\,+\,1}}S(\mu)
 \label{Deriv-S}
\end{equation}

By adopting the crossover EoS approach, the requirement of pressure equality between the hadronic and quark phases, which is characteristic of the Maxwell construction for a phase transition, is abandoned. Instead, the matter consists of simultaneous contributions from both hadronic and quark degrees of freedom. The relative contribution of each phase is prescribed by the switching function, which determines the corresponding weights in the pressure and is expressed as a function of the chemical potential $\mu$. Thus, the system is described by a single phase whose composition depends on the switching function and its parameters.

As can be clearly seen, the switching function in Eq.~(\ref{SF-1a}) depends on two free parameters, $\mu_0$ and $r$. The main objective of this subsection is to determine, or at least constrain, the values of these parameters. The parameters will be restricted to a narrow range to ensure that the crossover EoS adequately represents both hadronic and quark matter. Particular attention must be paid to the crossing point of the hadronic and quark EoSs and the region in its immediate vicinity. 

The crossover EoS should describe the transition from the hadronic phase to the quark phase as smoothly as possible. This is achieved by fixing $\mu_0$ to a value close to the chemical potential at the crossing point. Specifically, we consider $\mu_0 = \mu_{\rm cros} \pm 100 \, ({\rm MeV})$. A systematic analysis shows that the closer $\mu_0$ is to $\mu_{\rm cros}$, the more accurately the crossover EoS reproduces the hadronic and quark EoSs in the vicinity of the crossing point.

Regarding the parameter $r$, its integer value determines how rapidly the crossover EoS converges to the quark EoS beyond the crossing point. Higher values of $r$ lead to faster convergence. However, a threshold exists beyond which further increases in $r$ have a negligible effect on the behavior of the crossover EoS. In the present study, satisfactory convergence is achieved for integer values in the range $r \in [6,9]$, i.e., $r \in \mathbb{Z}$.

\begin{figure}[ht]
    \centering
    \includegraphics[width=1.0\linewidth]{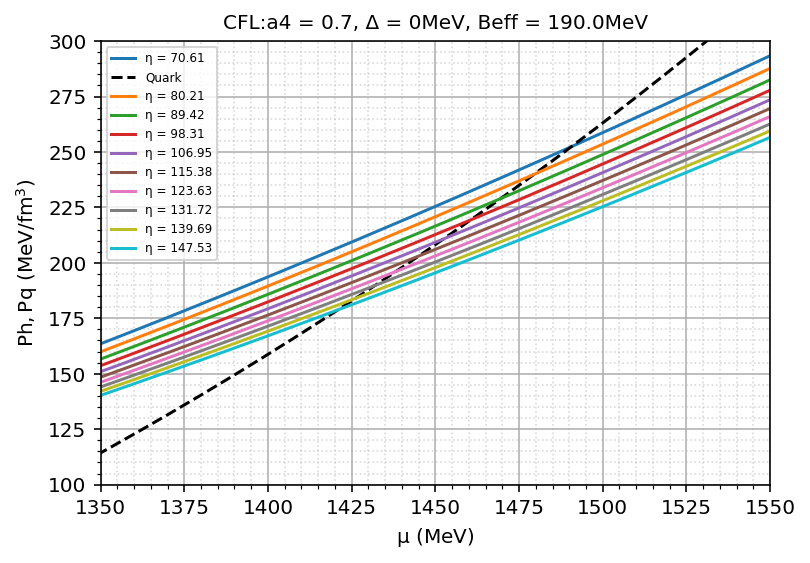}
    \caption{Pressure as a function of the chemical potential for the hadronic EoSs (colored lines) and the quark EoS (black dashed line). The figure illustrates the crossing points and the corresponding pressures and chemical potentials at which each hadronic EoS intersects the quark EoS.}
    \label{fig:crossing points}
\end{figure}

The following figures illustrate the aforementioned behavior of the crossover EoS as a function of the switching function and its free parameters. Thus, as shown in Fig.~\ref{fig:crossing points}, an appropriate choice of hadronic and quark EoSs yields a crossing point that marks the transition between the two phases. The main objective is to construct a crossover EoS that adequately describes hadronic matter below the crossing point, quark matter above it, and the smooth transition between the two in its vicinity. To achieve this, the parameters are chosen as follows: $\mu_0\,=\,\mu_{\rm cros}\,\pm\,100\,{\rm MeV} $ and $r \in [6,9]\,\cap \,\&\,r\in \mathbb{Z}$

\begin{figure}[ht]
    \centering
    \includegraphics[width=1.0\linewidth]
    {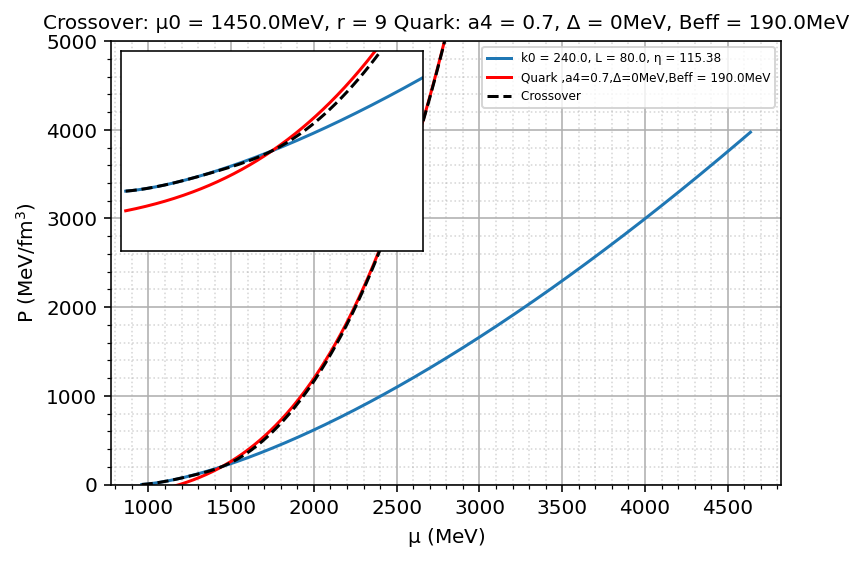}
    \caption{Pressure as a function of the chemical potential for the hadronic (blue line), quark (red line), and crossover (black dashed line) EoSs. By choosing $\mu_0$ close to the chemical potential at the crossing point and setting $r \in [6,9] \cap \mathbb{Z}$, the crossover EoS adequately reproduces the hadronic EoS below the crossing point, the quark EoS above it, and the behavior of both EoSs in the vicinity of the crossing point.}
    \label{fig:Crossover1}
\end{figure}

\begin{figure}[ht]
\centering
\includegraphics[width=1.0\linewidth]{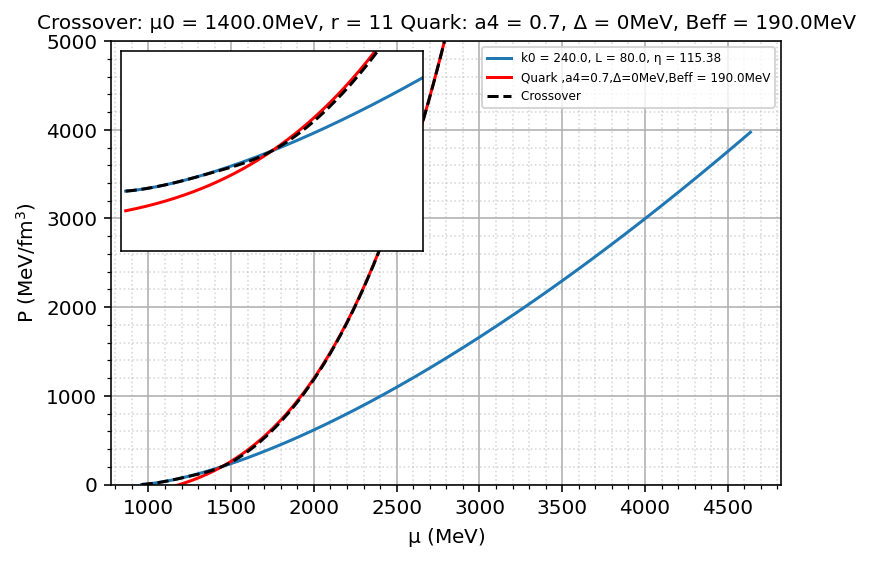}
\caption{Pressure as a function of the chemical potential for the hadronic (blue line), quark (red line), and crossover (black dashed line) EoSs. By choosing $\mu_0$ close to the chemical potential at the crossing point and setting $r \in [6,9] \cap \mathbb{Z}$, the crossover EoS adequately reproduces the hadronic EoS below the crossing point and the quark EoS above it, while also capturing their behavior in the vicinity of the crossing point.}
    \label{fig:Crossover2}
\end{figure}

As shown in Fig.~\ref{fig:Crossover1}, by choosing $\mu$ and $r$ according to the aforementioned criteria, the crossover EoS reproduces the hadronic EoS remarkably well below the crossing point and the quark EoS above it. It also provides a satisfactory description of the transition in the vicinity of the crossing point. The same desirable behavior is observed when $\mu_0$ is varied slightly while remaining close to $\mu_{\rm cros}$ and the value of $r$ is increased. As demonstrated in Fig.~\ref{fig:Crossover1} and Fig.~\ref{fig:Crossover2} choosing $r \geq 6$ and $\mu_0 = \mu_{\rm cros} \pm 100\,{\rm MeV}$ allows the crossover EoS to reproduce both EoSs satisfactorily below, above, and in the vicinity of the crossing point. 

On the other hand, it is clear that choosing values of $\mu_0$ that deviate significantly from $\mu_{\rm cros}$, or setting $r<6$ compromises the ability of the crossover EoS to adequately reproduce both phases and their behavior in the vicinity of the crossing point.

\begin{figure*}[t]
\centering
\includegraphics[width=0.47\textwidth]
{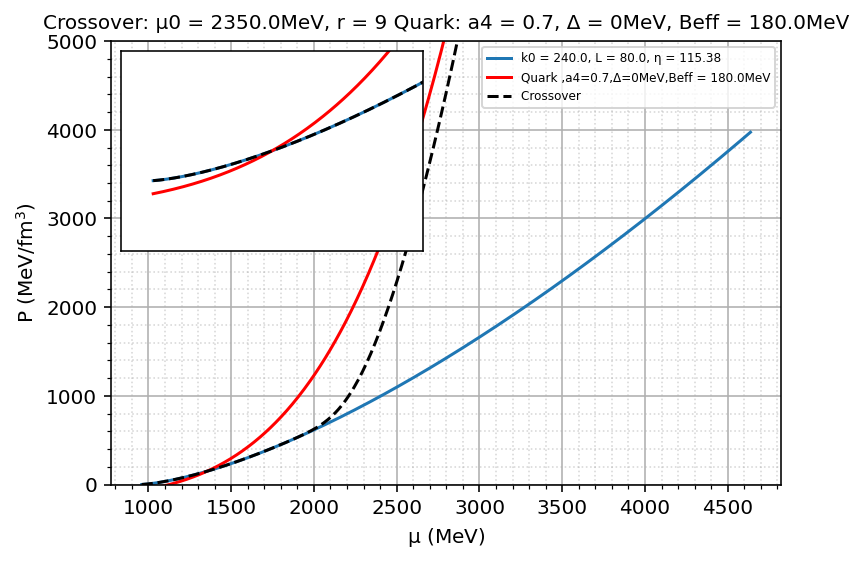}
\includegraphics[width=0.47\textwidth]
{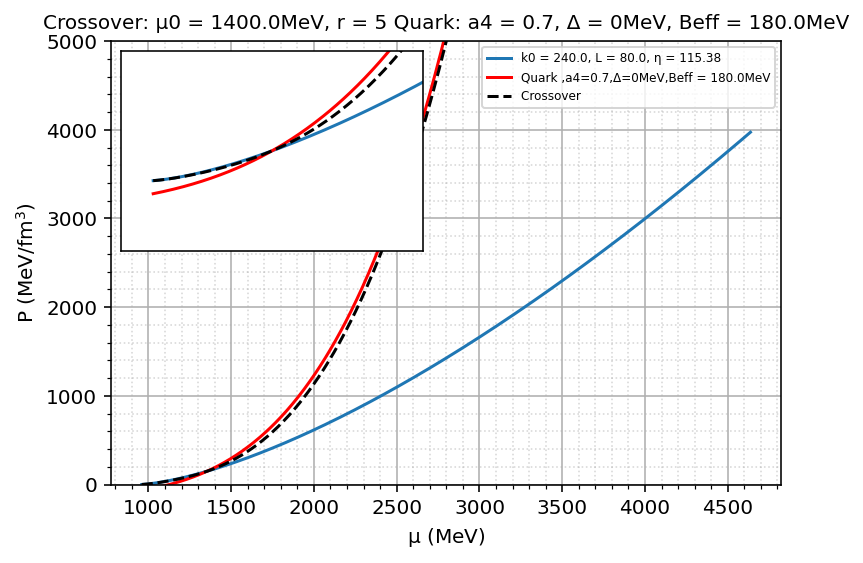}
\caption{Pressure as a function of the chemical potential for the hadronic (blue line), quark (red line), and crossover (black dashed line) EoSs. Choosing $\mu_0$ far from the chemical potential at the crossing point (left panel) or setting $r \leq 5$, with $r \in \mathbb{Z}$ (right panel), results in a poor representation of the quark EoS beyond the crossing point and of the behavior of the EoSs in the vicinity of the crossing point.}
    \label{fig:ngcrossover}
\end{figure*}

As shown in Fig.~\ref{fig:ngcrossover}, the inability of the crossover EoS to adequately represent hadronic matter, quark matter, and the region around the crossing point is evident. In the left panel, $\mu_0$ deviates significantly from $\mu_{\rm cros}$, resulting in an unsatisfactory description of quark matter by the crossover EoS. Although the crossover EoS follows the hadronic EoS, it fails to converge to the quark EoS beyond the crossing point. This is an undesirable outcome. In the right panel, although $\mu_0$ remains close to $\mu_{\rm cros}$, the very low integer value of $r$ leads to slow convergence of the crossover EoS toward the quark EoS beyond the crossing point. In other words, the crossover construction fails to reproduce the behavior of the quark EoS adequately.

\section{Neutron star structure}
The bulk properties of neutron stars, such as mass, radius and tidal deformability are determined by solving the coupled TOV  equations~\cite{{Glendenning1996CompactSN},Shapiro:1983du,Haensel2007NeutronS1,schaffner-bielich_2020}.  As discussed in the preceding section, the crossover equation of state ${\cal E}={\cal E}(P)$ serving as the fundamental input to the TOV equations. 
This equations read
\begin{eqnarray}
\frac{dP(r)}{dr}&=&-\frac{G{\cal E}(r) M(r)}{c^2r^2}\left(1+\frac{P(r)}{{\cal E}(r)}\right) \nonumber \\
&\times&
 \left(1+\frac{4\pi P(r) r^3}{M(r)c^2}\right) \left(1-\frac{2GM(r)}{c^2r}\right)^{-1},
\label{TOV-1}
\end{eqnarray}
\begin{equation}
\frac{dM(r)}{dr}=\frac{4\pi r^2}{c^2}{\cal E}(r)
\label{TOV-2}
\end{equation}
It is worth noting here that in recent years, valuable insights have been gained from observing gravitational waves produced by the mergers of black hole–neutron star and neutron star–neutron star binary systems. These events provide an opportunity to measure various properties of neutron stars. Notably, during the inspiral phase of binary neutron star systems, tidal effects can be detected. More specifically, the tidal Love number $k_2$ 
characterizes how a neutron star responds to an external tidal field $E_{ij}$, depending on both its mass and the applied equation of state. In particular,  the induced quadrupole moment $Q_{ij}$ due 
to the applied tidal $E_{ij}$ is given by~\cite{Flanagan-2008,Hinderer-2008}
\begin{equation}
Q_{ij}=- \lambda E_{ij},
\label{Love-1}
\end{equation}
where  $\lambda=2k_2R^5/3G$ is the tidal deformability. 
In addition, an important and well measured quantity by the gravitational wave detectors, which can be treated as a tool to impose constraints on the EoS, is the dimensionless tidal deformability $\Lambda$, defined as 
\begin{equation}
    \Lambda=\frac{2}{3}k_2 \left(\frac{c^2R}{GM}\right)^5=\frac{2}{3}k_2 (1.473)^{-5}\left( \frac{R}{{\rm Km}} \right)^5\left(\frac{M_{\odot}}{M}  \right)^5
\label{dimens-tidal}    
\end{equation}
We notice that $\Lambda$ is sensitive to the neutron star radius, hence can provide information for the low-density part of the EoS, which is also related to the structure and properties of finite nuclei. Consequently, since in some cases the crossover point may occur at relatively low densities, it is of particular interest to investigate the effect of the phase transition on the neutron-star radius and, consequently, on the tidal deformability, which is, in principle, an observable quantity. The correlation between the value of $\Lambda$ and the crossover point may potentially provide valuable information about the properties of matter in the corresponding density regime.

\section{Results and Discussion}
In this section the main results obtained using the crossover EoSs are presented. Specifically, ten hadronic EoS, corresponding to ten different values of the parameter $\eta$, are combined with a single quark EoS based on the model presented in Ref.~\cite{BlaschkeHanu-2022}. A range of quark EoSs is obtained by varying the relevant model parameters, $a_4$, $\Delta$ and $B_{\rm eff}$ (see Table~\ref{Table-2}).

\subsection{Crossing point and transition from hadrons to quarks }
\begin{figure}[ht]
\centering
\includegraphics[width=1.0\linewidth]
{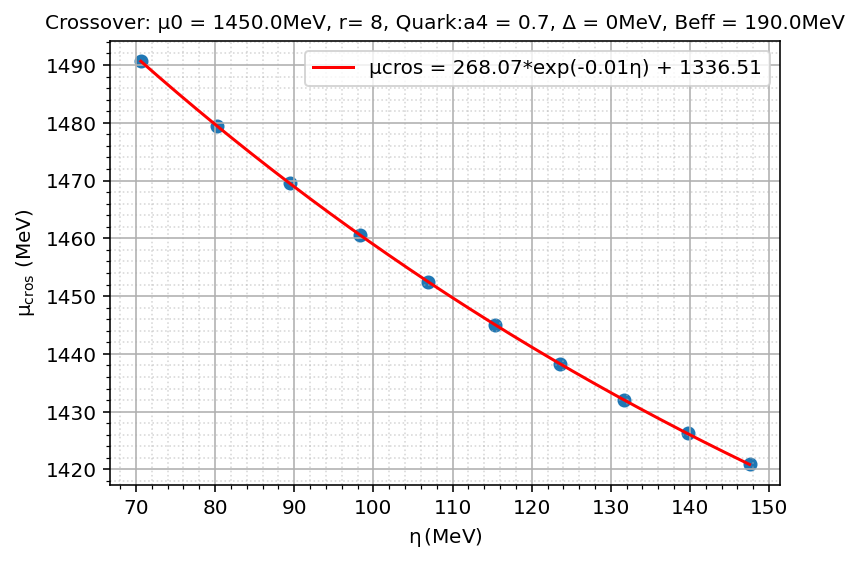}
\caption{Chemical potential at the crossing point as a function of the parameter $\eta$ for each hadronic EoS combined with a given quark EoS.}
\label{fig:mcrosh}
\end{figure}

To begin with we study the pressure of hadronic matter as a function of the chemical potential for the selected values of the parameter $\eta$. In each case, we construct the corresponding equation of state for quark matter with a specific parametrization and investigate the resulting crossing point. In all cases, the stiffer the hadronic equation of state, the more the crossing point with the corresponding quark-matter equation of state shifts toward lower values of the chemical potential. Clearly, the location of the crossing point depends both on the values of the parameter $\eta$ characterizing the hadronic equation of state and on the parametrization adopted for the quark-matter equation of state. 
To illustrate the dependence of the chemical potential at the crossing point on the parameter ($\eta$) characterizing the hadronic equation of state, we present Fig.~\ref{fig:mcrosh}. For a specific parametrization of the CFL equation of state, we obtain a linear dependence of the form
\begin{equation}
\mu_{\rm cros}=p_1 \exp[-p_2 \eta]+p_3
\label{fit-1}
\end{equation}
where the parameters $p_1$, $p_2$ and $p_3$ depend on the specific parametrization of the quark equation of state. For example in Fig. \ref{fig:mcrosh} the corresponding values of the parametres are $p_1=268.07$, $p_2=0.01$ and $p_3=1336.51$. 

\begin{figure*}[t]
    \centering
    \includegraphics[width=0.46\textwidth]{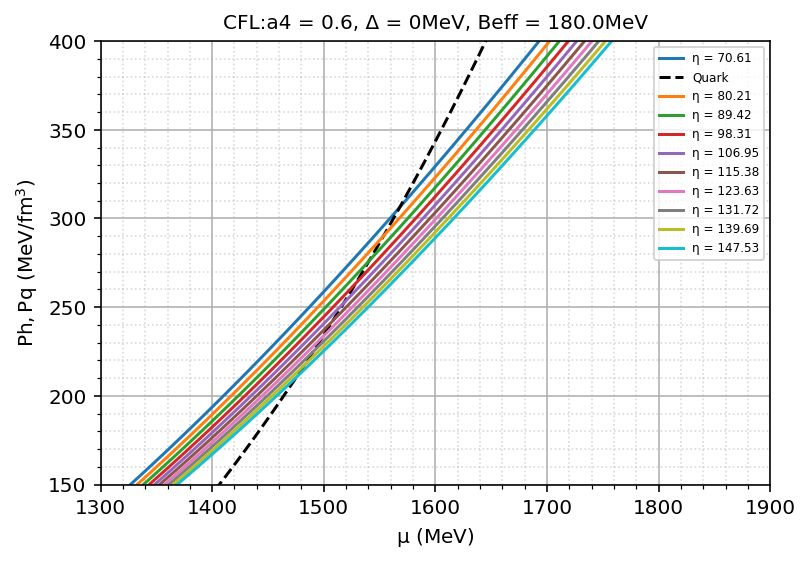}
    \hfill
    \includegraphics[width=0.46\textwidth]{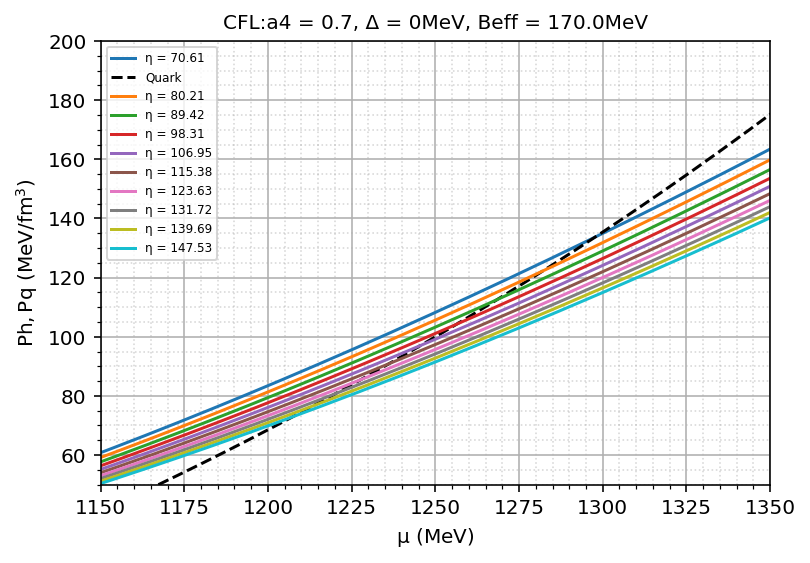}
    \hfill
    \includegraphics[width=0.46\textwidth]{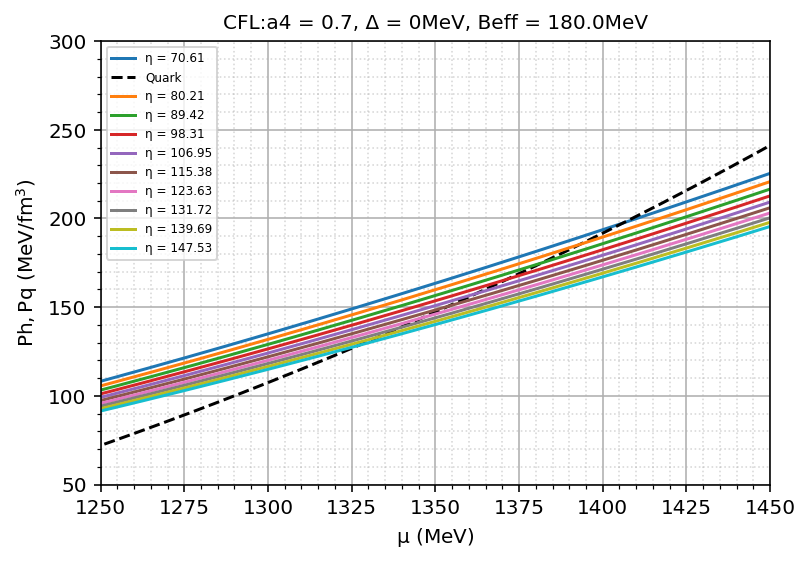}
\hfill
    \includegraphics[width=0.46\textwidth]{FPm070190b.png}
    \hfill
    \includegraphics[width=0.46\textwidth]{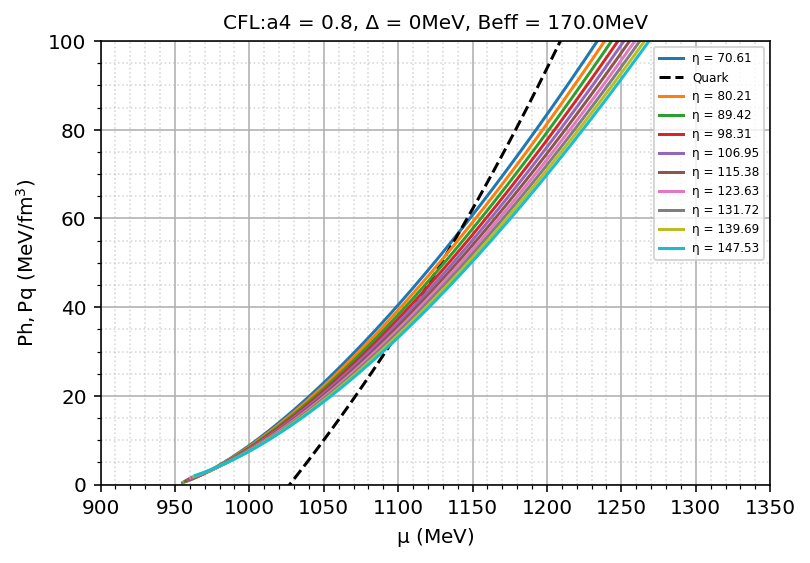}
    \hfill
   \includegraphics[width=0.46\textwidth]{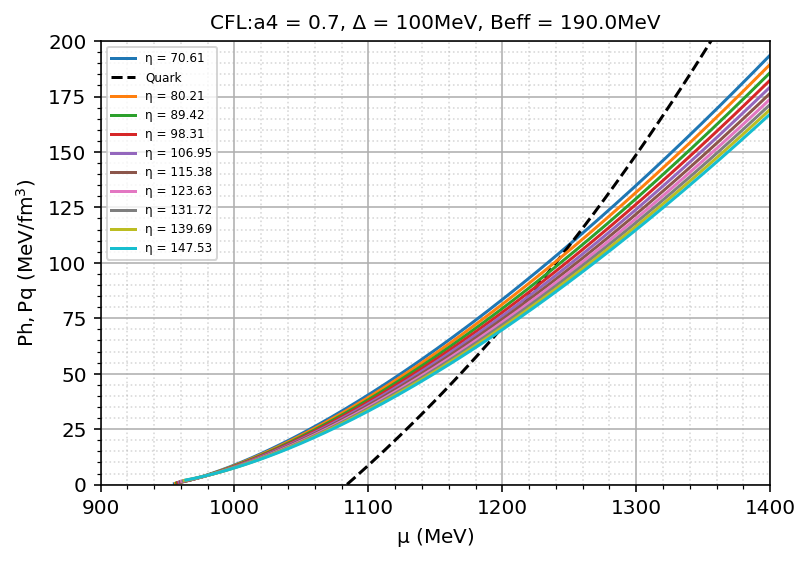}
\caption{Pressure as a function of the chemical potential for the hadronic EoSs (colored solid lines) and the quark EoS (black dashed line). Setting $\Delta = 0\,{\rm MeV}$ and decreasing $a_4$ shifts the crossing points between the hadronic and quark EoSs toward higher pressures and higher chemical potentials. The parameter $B_{\rm eff}$ has a smaller influence on the crossing points, with higher values of $B_{\rm eff}$ leading to higher crossing pressures and corresponding chemical potentials. Setting $\Delta = 100\,{\rm MeV}$ (bottom right panel), requires either higher values of $B_{\rm eff}$ or lower values of $a_4$ to avoid crossing points at excessively low chemical potentials and pressures, as well as cases in which no crossing point occurs. }
    \label{fig:Pm crossing point a4}
\end{figure*}

Additionally, the location of the crossing point should be regarded as a multiparameter problem. For example, the lower the value of $a_4$, the higher the crossing point occurs in terms of both pressure and chemical potential, as shown in Fig.~\ref{fig:Pm crossing point a4}. Moreover, as the value of $B_{\rm eff}$ increases, the crossing point is shifted toward higher values of both pressure and chemical potential, as illustrated in Fig.~\ref{fig:Pm crossing point a4}.


Changing the value of $\Delta$ affects the pressure and the chemical potential at which the crossing point occurs. Specifically, higher values of $\Delta$ shift the crossing point toward lower chemical potentials and pressures. For this reason, we set $\Delta = 0\, {\rm MeV}$. Increasing $\Delta$ may result in crossing points at very low chemical potentials and pressures or may eliminate the crossing point altogether. In other words, choosing $\Delta = 100\, {\rm MeV}$ requires either lower values of $a_4$ or substantially higher values of $B_{\rm eff}$. Table~\ref{Table-2} lists the value $\Delta = 100\,{\rm MeV}$ and the corresponding values of $a_4$ and $B_{\rm eff}$ that yield a crossing between the hadronic and quark EoSs. For these parameter values, the crossing occurs at relatively low chemical potentials and pressures, as can be seen in Fig.~\ref{fig:Pm crossing point a4}.

\begin{figure}[ht]
\centering
\includegraphics[width=1.0\linewidth]
{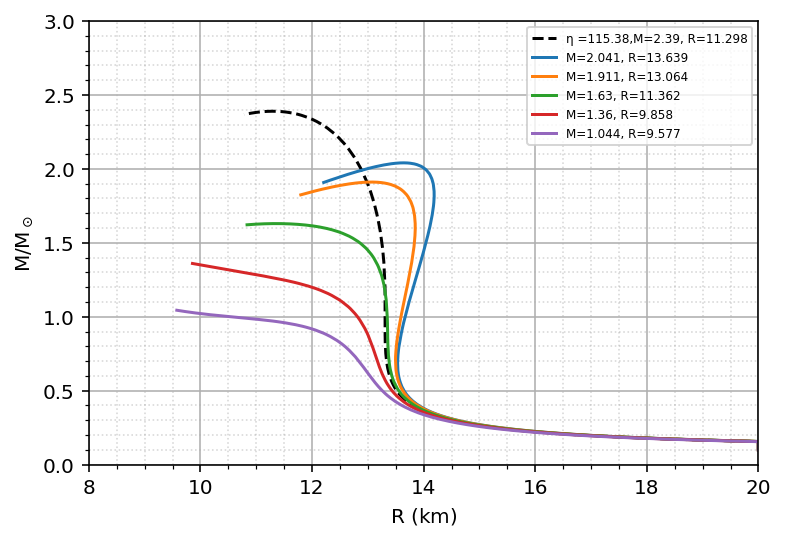}
    \caption{Mass–radius diagram for five quark EoSs (colored lines) corresponding to different values of $\Delta$, with $a_4 = 0.7$ and $B_{\rm eff} = 190\,{\rm MeV}$, and a crossover EoS (black dashed line) corresponding to $\eta = 115.38$ MeV. }
    \label{fig:Various Delta MR}
\end{figure}

\begin{figure}[ht]
    \centering
\includegraphics[width=1.0\linewidth]
{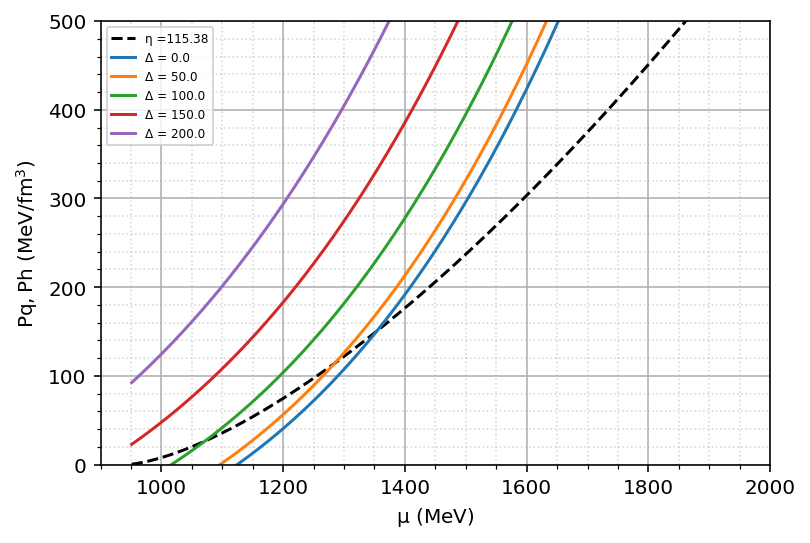}
\caption{Five quark EoSs (colored lines) for different values of $\Delta$, with $a_4 = 0.7$ and $B_{\rm eff} = 190\,{\rm MeV}$, compared with a hadronic EoS (black dashed line) corresponding to $\eta = 115.38$ MeV. }
    \label{fig:Various Delta}
\end{figure}

It is clear that, for this specific set of parameters, not all values of $\Delta$ yield a crossing point. This may lead to cases that are of no physical interest, at least within the framework of the present crossover model. Specifically, a model without a crossing point may be mathematically or computationally acceptable, but it cannot be regarded as a genuine crossover model. In other words in Fig.~\ref{fig:Various Delta MR}, the purple and red curves, although mathematically well-defined and apparently reasonable, do not correspond to a genuine crossover within the framework of the present model. This is confirmed by Fig.~\ref{fig:Various Delta}, which shows that the quark EoSs corresponding to $\Delta = 150\,{\rm MeV}$ (red curve) and $\Delta = 200\,{\rm MeV}$ (purple curve) do not intersect the hadronic EoS.


\subsection{Equations of state, speed of sound and bulk neutron stars properties}
\begin{figure*}[t]
\centering
\includegraphics[width=0.45\textwidth]{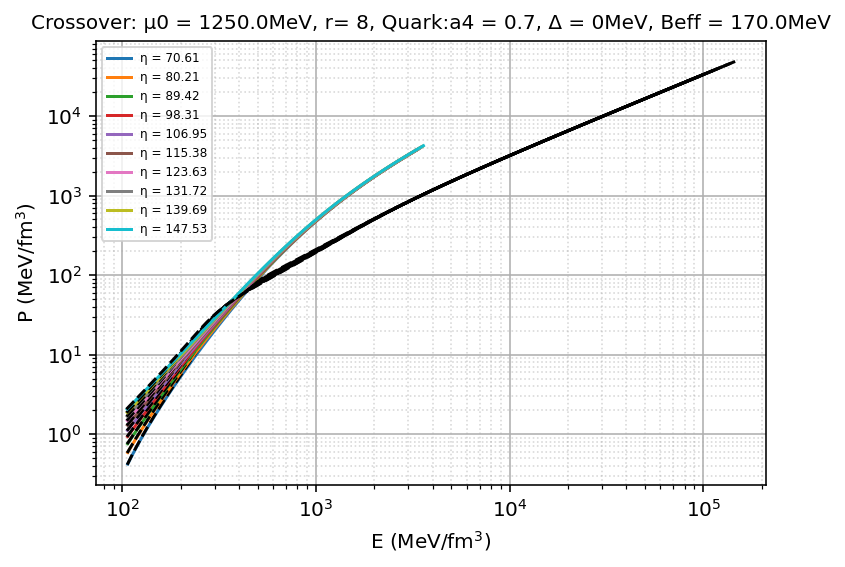}
    \hfill
\includegraphics[width=0.45\textwidth]{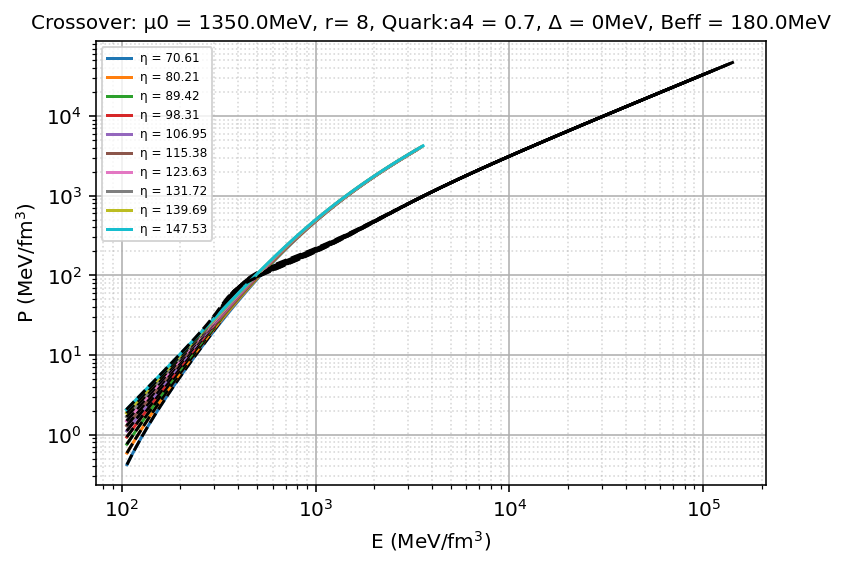}
    \hfill
\includegraphics[width=0.45\textwidth]{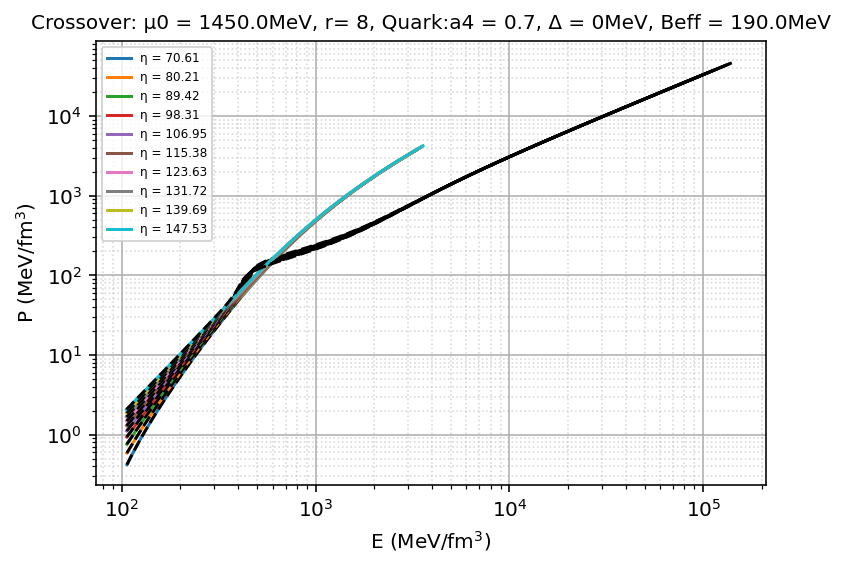}
    \hfill
\includegraphics[width=0.45\textwidth]{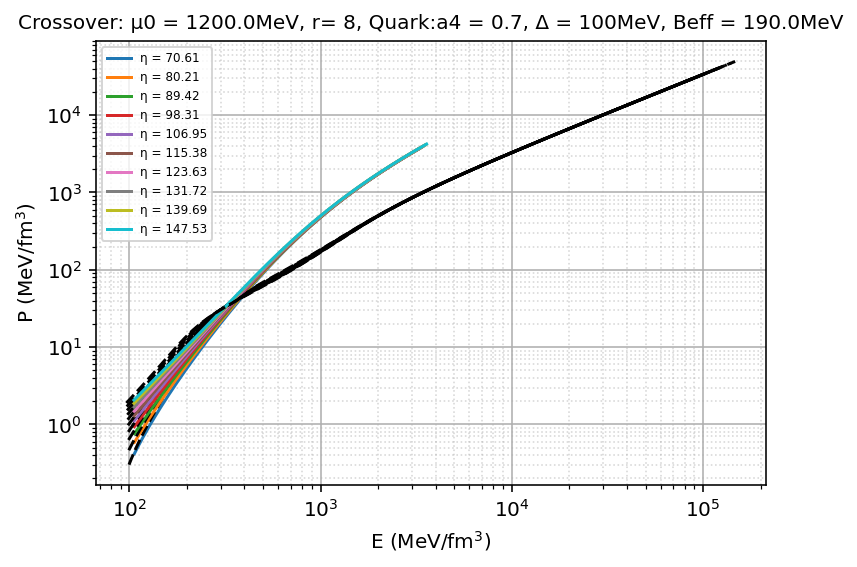}
\caption{Pressure as a function of energy density for the hadronic EoSs (colored lines) and the corresponding crossover EoSs (black dashed lines). Below the crossing point, the crossover EoSs closely reproduce the hadronic EoSs. The transition around the crossing point is smooth. Beyond the crossing point, the inclusion of quark matter softens the crossover EoSs (black dashed lines) relative to the corresponding hadronic EoSs (colored lines).}
    \label{fig:PcEc diagram}
\end{figure*}

The hybrid equations of state are constructed using the switching-function method described in Section II. In each case, an appropriate choice of the parameters $r$ and $\mu_0$ is required. The results are presented in Fig.~\ref{fig:PcEc diagram}, where both the hadronic equations of state and the corresponding hybrid equations of state are shown for comparison. 

It is important to emphasize that the parametrization of the switching function is not arbitrary. Instead, the value of the parameter $\mu_0$ has been chosen to be close to the chemical potential at which the crossing occurs, within a range of $\pm 100\,{\rm MeV}$. At the same time, the parameter $r$ has been selected so as to ensure the smoothest possible transition from hadronic to quark matter, thereby guaranteeing a smooth phase transition. Thus, within this approach, the parameter $\mu_0$ is not only assigned a clear physical meaning, but the resulting phase transition is also ensured to be smooth.
The switching function, as was defined in Eq.~\eqref{SF-1a}, is governed by two parameters $\mu_0$ and $r$. Varying these parameters, crossing point, namely the pressure and the chemical potential where the transition from hadrons to quarks occurs, remains intact. This is because the EoS for hadronic and for quark matter are not affected by any of these two parameters. However, changes to these parameters have significant impact on crossover EoS and on the fashion that represents hadronic and quark EoS before and after crossing point, respectively. Additionally, the values of these parameters affect the behavior of crossover EoS on the crossing point and its vicinity.

The main goal of constructing a crossover EoS is to represent properly hadronic and quark phases, as well as, their crossing point. In order to achieve this the integer value of $r$ has to be $r \in \mathbb{Z}\, \cap\,r \in [6,9]$. This parameter has to be in this range to contribute to an efficient crossover EoS. It is obvious that values of $r$ out this range lead crossover EoS to diverge from quark matter after the crossing point.

Additionally, parameter $\mu_0$ has to serve the same goal as $r$, namely, an efficient crossover EoS. In order to accomplish that $\mu_0\,=\,\mu_{\rm cros}\,\pm\,100$. This parameter has great impact on crossover's behavior and its ability to represent efficiently both EoSs. Values of $\mu_0$ far from $\mu_{\rm cros}$ lead to a totally unacceptable behavior for crossover EoS.

\begin{figure*}[!t]
\centering
\includegraphics[width=0.43\textwidth]{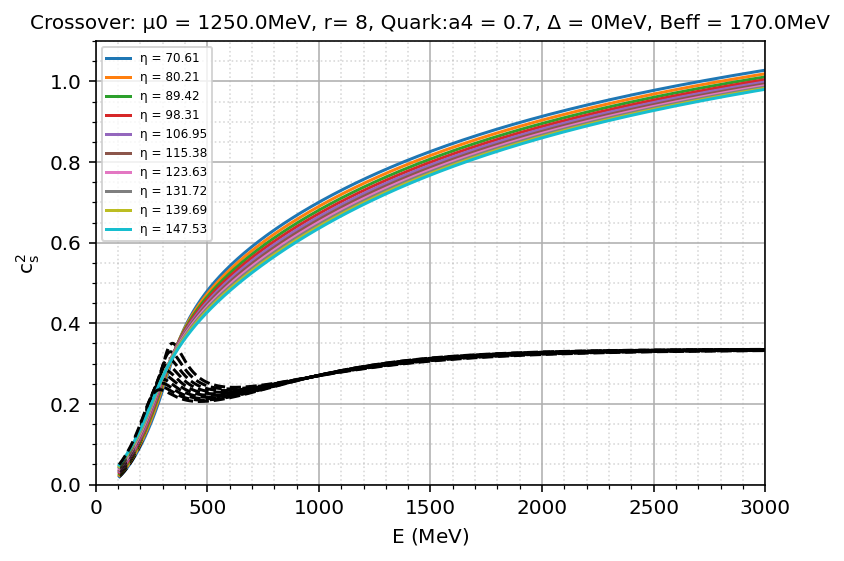}
\hfill
\includegraphics[width=0.43\textwidth]{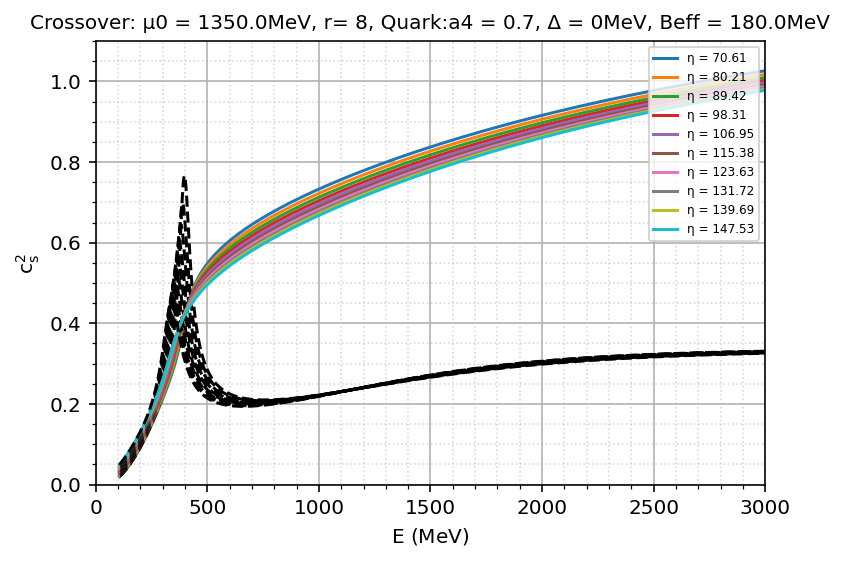}
\hfill
\includegraphics[width=0.43\textwidth]{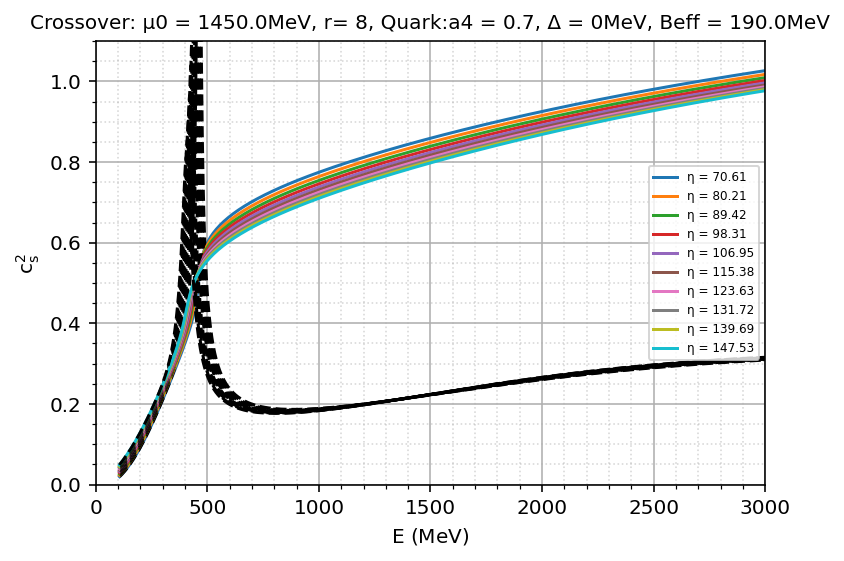}
\hfill
\includegraphics[width=0.43\linewidth]{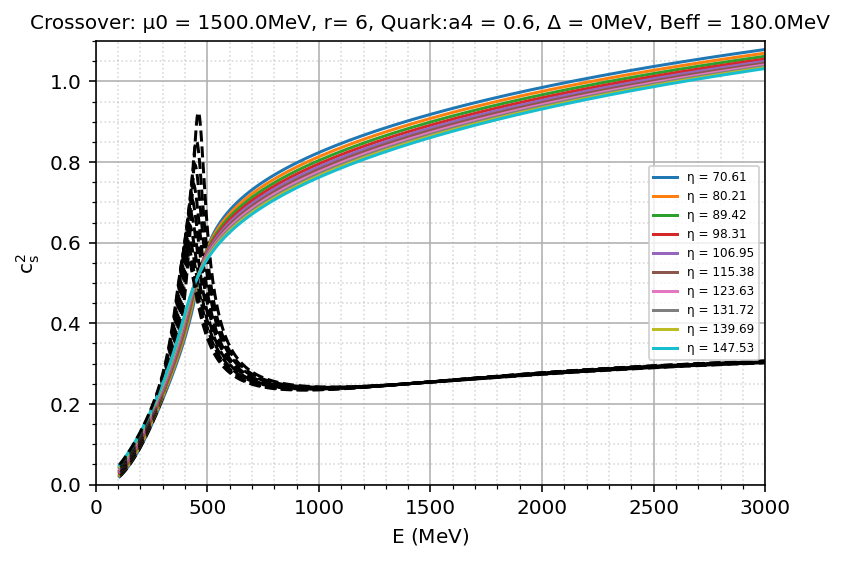}
\hfill
\includegraphics[width=0.43\linewidth]{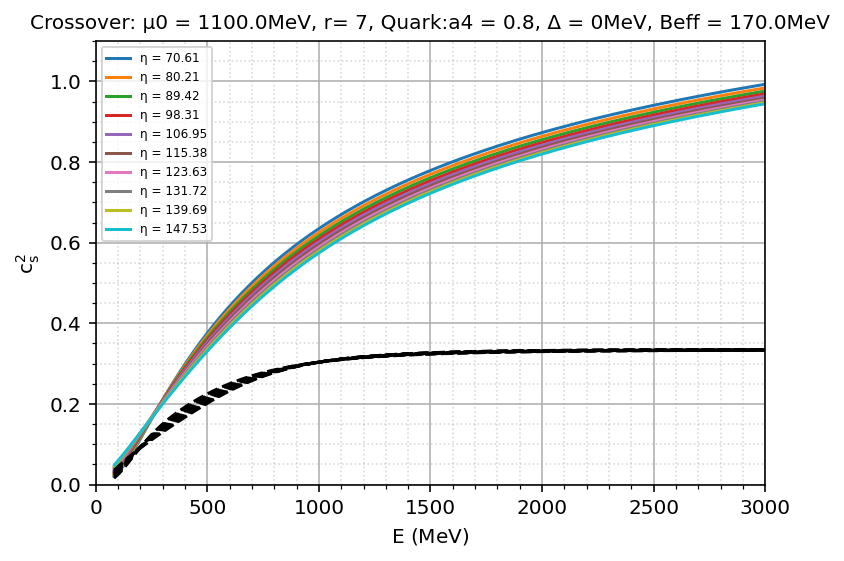}
\hfill
\includegraphics[width=0.43\linewidth]{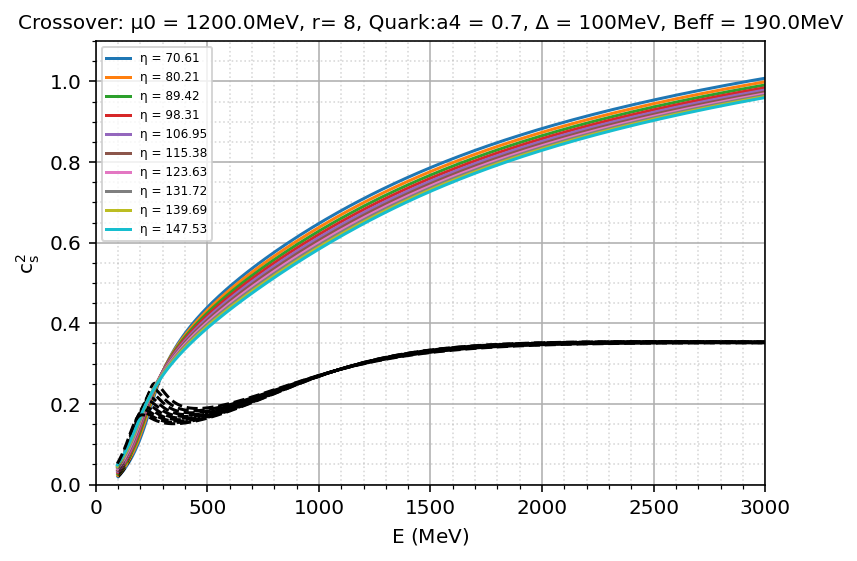}
\caption{Squared speed of sound as a function of energy density for hadronic matter (colored lines) and the corresponding crossover EoSs (black dashed lines). The inclusion of quark matter in the crossover EoSs produces a kink in the squared speed of sound, which subsequently approaches the value of $1/3$. Beyond the crossing point, the squared speed of sound for the crossover EoSs deviates from that of the hadronic EoSs. Higher values of $B_{\rm eff}$ result in a more pronounced peak in the vicinity of the crossing point.}
    \label{fig:cs2 diagram}
\end{figure*}


\begin{figure*}[t]
\centering
\includegraphics[width=0.45\textwidth]{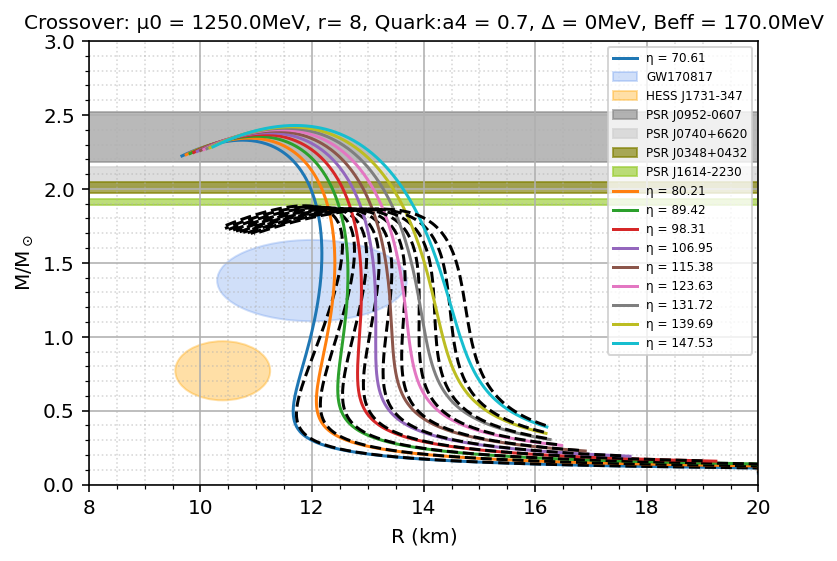}
    \hfill
\includegraphics[width=0.45\textwidth]{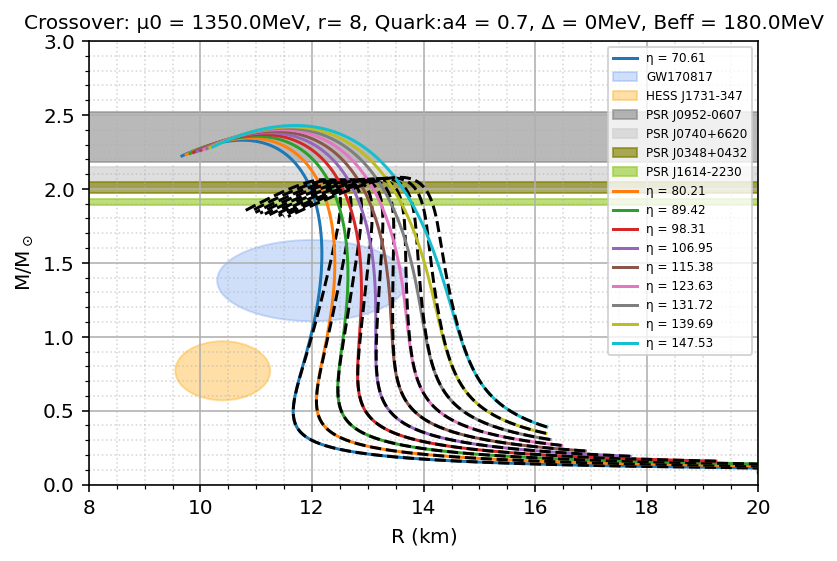}
    \hfill
\includegraphics[width=0.45\textwidth]{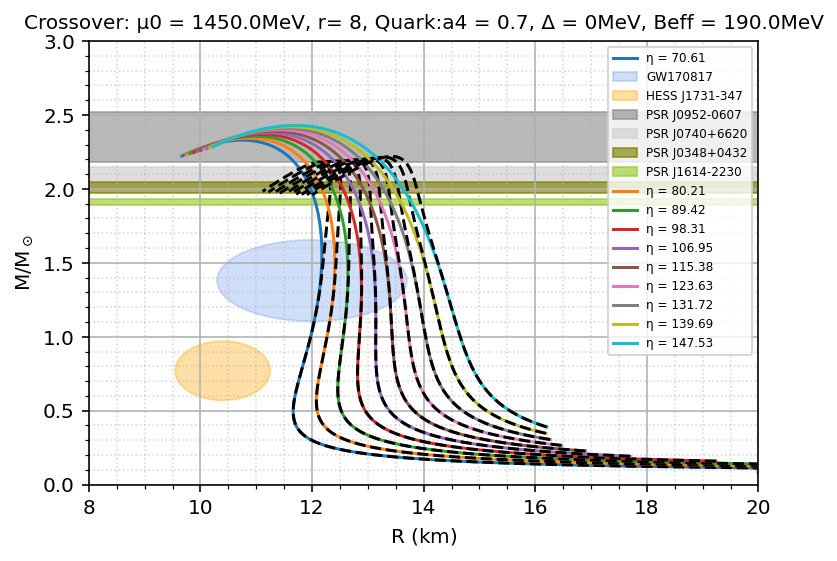}
\hfill
\includegraphics[width=0.45\linewidth]{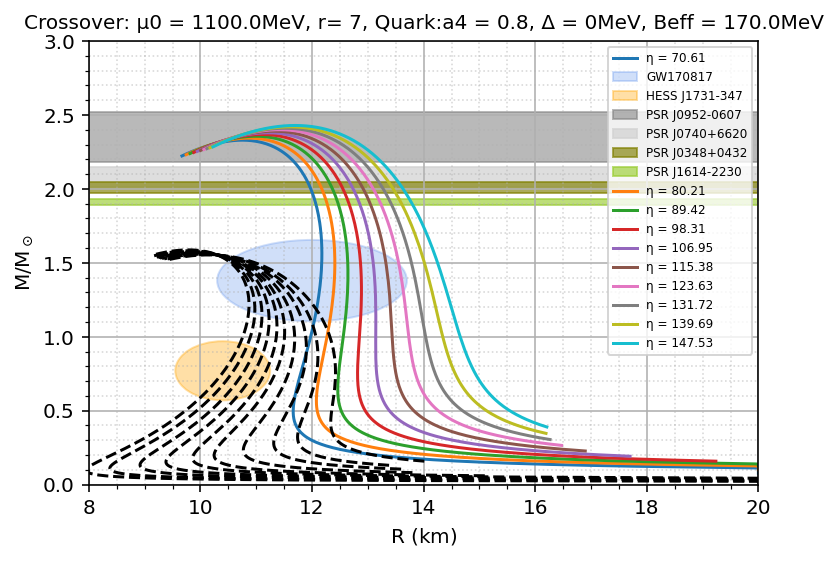}
\includegraphics[width=0.45\linewidth]{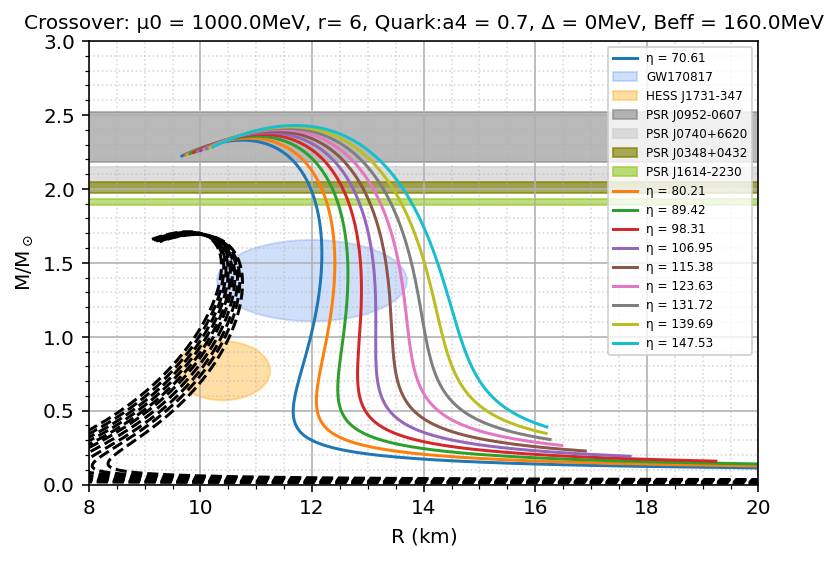}  
\hfill  
\includegraphics[width=0.45\linewidth]{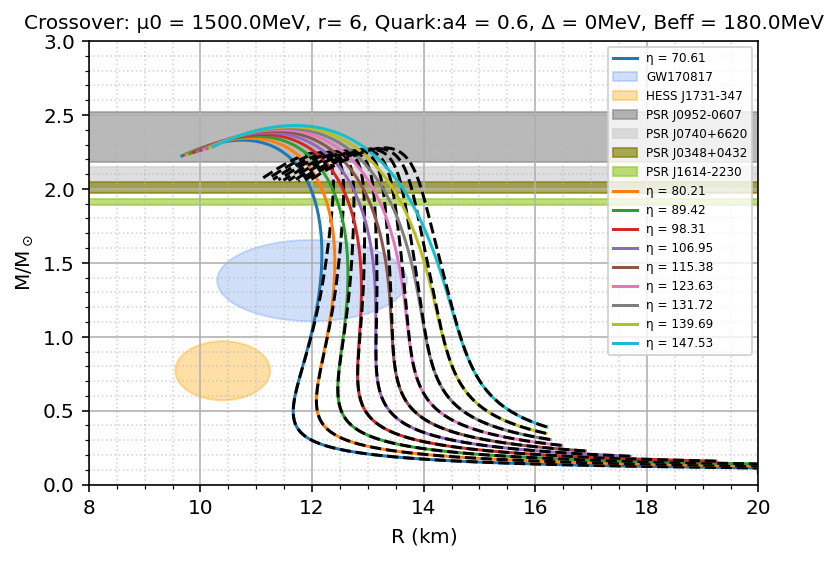}
\hfill  
\includegraphics[width=0.45\linewidth]{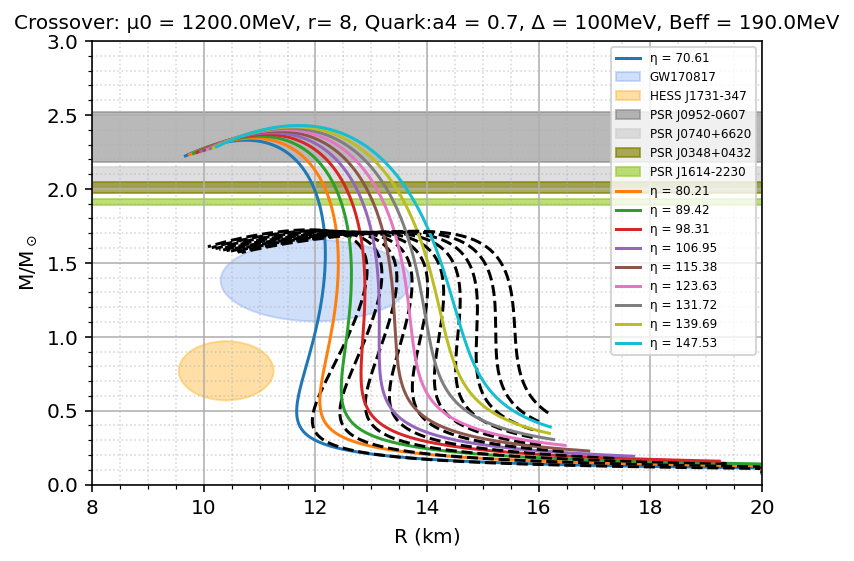}
\caption{Mass–radius diagrams for crossover EoSs, with $\mu_0$ and $r$ chosen according to the crossing-point values. The effect of varying $B_{\rm eff}$ on the maximum mass and corresponding radius is investigated for $a_4 = 0.7$ and $\Delta = 0\,{\rm MeV}$. Higher values of $B_{\rm eff}$ lead to higher maximum masses and corresponding radii. For $B_{\rm eff} \geq 180\,{\rm MeV}$, the maximum mass exceeds two solar masses. The bottom panel shows the corresponding results for $\Delta = 100\,{\rm MeV}$. The shaded
regions from bottom to top represent the HESS J1731-347 remnant
\cite{Doroshenko-2022}, the GW170817 event \cite{Abbott-2019-X} as well as the PSR J1614-2230 \cite{Arzoumanian-2018}, PSR
J0348+0432 \cite{Antoniadis-2013}, PSR J0740+6620 \cite{Cromartie-2020}, and PSR J0952-0607  
\cite{Romani-2022} pulsar observations for the possible maximum mass.  }
    \label{fig:MR diagramBeff}
\end{figure*}

The most striking feature of the figures presented above is that, as the hadronic equation of state becomes stiffer, the crossing point occurs at lower densities, implying an earlier onset of quark matter. Consequently, the corresponding maximum masses are also reduced compared with those obtained for the purely hadronic equations of state. This is a well-known characteristic of hybrid equations of state in general. It is evident that, in some cases, the hybrid equations of state must be ruled out because they are unable to support neutron-star masses (or, more generally, compact-object masses) above two solar masses, as required by the existing observational evidence. In all cases, an appropriate combination of the quark-matter parameters is required in order to ensure that the resulting predictions are consistent with the available observations. In principle, one could perform a systematic investigation, possibly with the aid of statistical methods, to determine the range of CFL-matter parameters that is compatible with observational constraints. However, this is beyond the scope of the present study.

It is therefore of particular interest to investigate the behavior of the speed of sound in the EoS, both in the presence and in the absence of the quark phase. This behavior is displayed in Fig.~\ref{fig:cs2 diagram}.
For purely hadronic matter, the squared speed of sound asymptotically approaches unity. The squared speed of sound in the crossover EoS follows the same trend as long as hadronic matter dominates. However, as the contribution of quark matter becomes significant, the squared speed of sound exhibits behavior characteristic of quark matter, approaching the value of $1/3$ in accordance with Eq.~(\ref{CFL-speed}). Overall, the squared speed of sound develops a peak before approaching $1/3$ as quark matter becomes dominant (see also Refs.~\cite{Rezzolla-2022a,Rezzolla-2022b}).

It is evident that the while the value of $B_{\rm eff}$ increases, the peak in square speed of sound for crossover EoS becomes higher and sharper, before the value of $1/3$ is approached. For crossing points at higher pressures and chemical potentials, the behavior characteristic of hadronic matter persists over a wider pressure range before the quark-matter phase becomes salient. 

On the other hand, for lower crossing points the quark matter and its behavior prevails quickly in low values for chemical potential and pressure.

The results for the corresponding mass–radius diagrams are presented in Fig.~\ref{fig:MR diagramBeff}. More specifically, we plot the hadronic equations of state together with their corresponding hybrid EoSs. In each case, the diagrams are constructed by considering a specific quark-matter EoS combined with the complete set of 10 selected hadronic EoSs.  
The first expected result is that hybrid equations of state are softer than their purely hadronic counterparts, generally predicting lower maximum masses. Clearly, the magnitude of this effect can be either reduced or enhanced by an appropriate choice of the quark equation-of-state parametrization. Of particular interest, however, is the case in which the quark equation of state is parametrized such that the crossover with the hadronic equation of state occurs at relatively low densities. Under these conditions, the influence of the quark equation of state, mediated by the switching function, may extend beyond the hadronic core, as expected, and affect the equation of state of the crust as well. This effect may be regarded as artificial, arising in part from the relatively low value of the chemical potential $\mu_0$, which, in the present approach, is chosen to be close to the crossover chemical potential $\mu_{\rm cross}$. 

An interesting implication of this approach is that, through an appropriate parametrization of the quark equation of state and the switching function, it becomes possible to predict exotic compact objects such as the HESS J1731–347 source, which cannot be reproduced using purely hadronic equations of state. In conclusion, within the present model, a phase transition occurring at relatively low densities, combined with a slowly varying switching function, may allow for the prediction of such objects with relatively low masses and small radii.

\begin{figure*}[t]
    \centering
\includegraphics[width=0.48\textwidth]{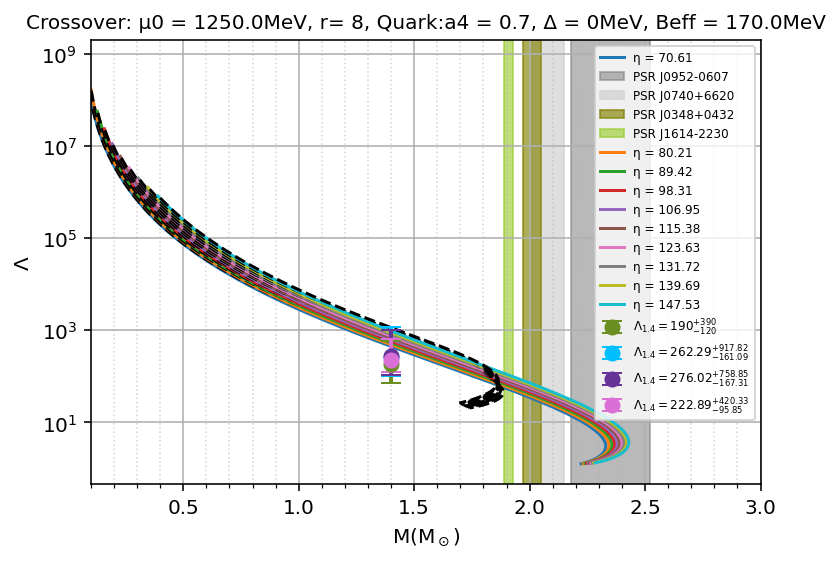}
    \hfill
\includegraphics[width=0.48\textwidth]{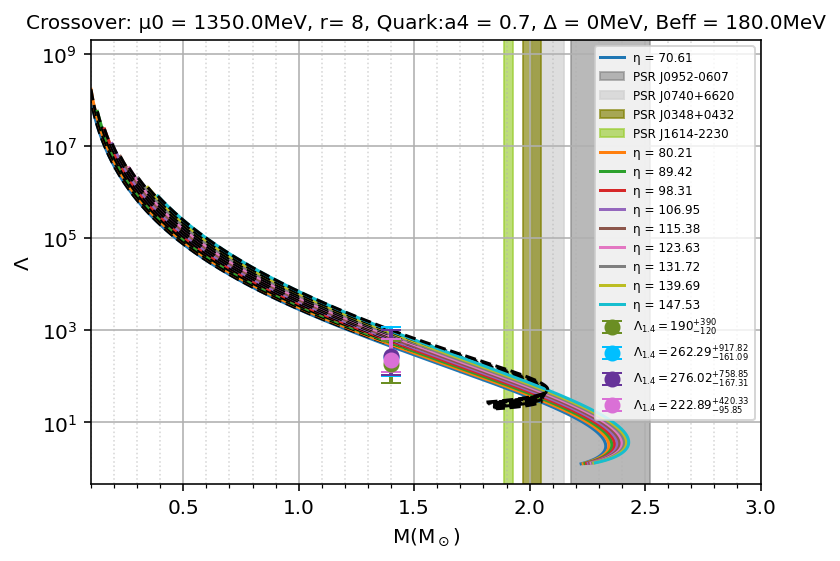}
    \hfill
\includegraphics[width=0.48\textwidth]{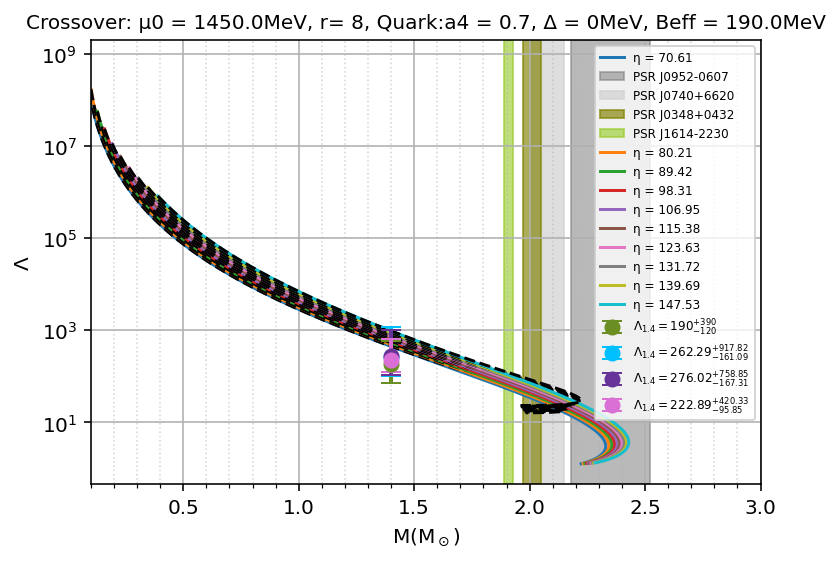}
    \hfill
\includegraphics[width=0.48\textwidth]{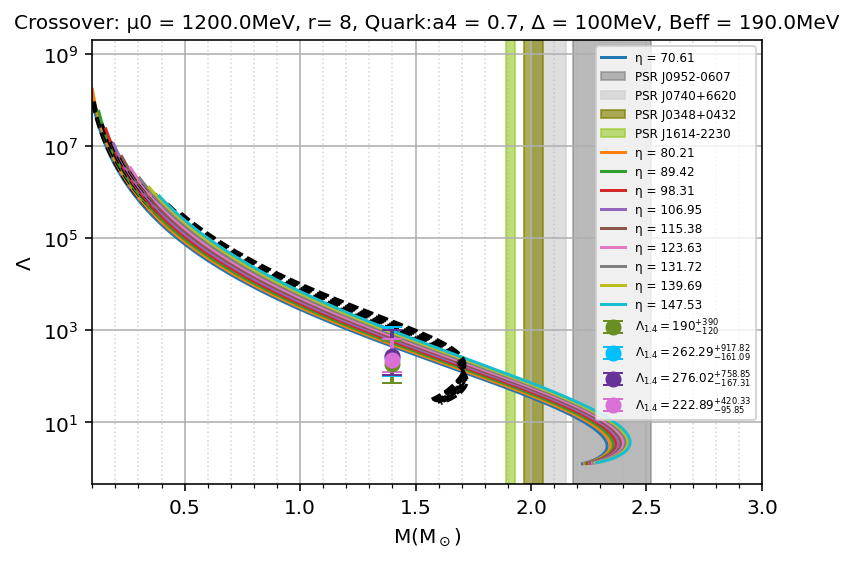} 
\caption{Tidal deformability $\Lambda$ as a function of mass. The inclusion of quark matter in the crossover EoSs (black dashed lines) reduces the maximum mass and the corresponding tidal deformability. Higher values of $B_{\rm eff}$ lead to higher maximum masses for the crossover EoSs. However, the maximum masses of the crossover EoSs remain lower than those of the corresponding hadronic EoSs. The observational data, shown with their corresponding error bars, indicate the estimated values of $\Lambda_{1.4}$ (corresponding to  mass $M=1.4 \ M_{\odot}$), provided by the GW170817 detection \cite{Abbott-2019-X} and also from the data-driven approach in Ref.~\cite{ Huang-2025}.}
\label{fig:LM diagram}
\end{figure*}

Finally, we examine the tidal deformability associated with the various crossover equations of state and by comparison with recent observational data~\cite{Abbott-2019-X} and results from   data-driven approach~\cite{ Huang-2025}. The results are presented in Fig.~\ref{fig:LM diagram}. As expected, the predicted tidal deformabilities are broadly similar to those obtained for the purely hadronic case. This behavior can be primarily attributed to the relatively small differences in stellar radii, particularly within the mass range encompassing the majority of observed neutron stars. Nevertheless, a more pronounced effect on the tidal deformability may be expected when the phase transition occurs at relatively low densities.
In this context, a systematic investigation of the tidal deformability of neutron stars with masses around $1.5 \ M{_\odot}$ could provide additional constraints on the density at which the hadron–quark phase transition takes place. Furthermore, such an analysis could help identify more realistic parameterizations of the switching function, thereby improving the description of the hadron–quark crossover.

\section{Concluding Remarks}

Overall, this study focuses on constructing an appropriate crossover equation of state  that consistently describes the hadronic phase, the quark phase, and the region around their crossing point. The hadronic EoS is controlled by the parameter $\eta=(K_0L^2)^{1/3}$, which depends on the symmetry energy at saturation density and its slope and serves as a regulator of the stiffness of the hadronic EoS. The quark EoS is determined by the parameters $a_4$, $\Delta$, and $B_{\rm eff}$. In the present study, we fix $a_4=0.7$ and  vary the pairing gap parameter $\Delta$ and also the effective bag constant $B_{\rm eff}$.

The two phases are connected through an exponential switching function, which provides a smooth crossover without introducing discontinuities or requiring an additional interpolation procedure. The parameters $\mu_0$ and $r$ of the switching function play a crucial role in ensuring that the resulting crossover EoS properly reproduces the behavior of both phases and, in particular, the vicinity of their crossing point. An efficient crossover construction requires $\mu_0$ to be sufficiently close to the chemical potential at the crossing, typically satisfying $\mu_0=\mu_{\rm cros}\pm100\, {\rm MeV}$, while the parameter $r$ is restricted to integer values in the range $6\leq r\leq9$. Within these constraints, the crossover EoS follows the hadronic EoS below the crossing region and gradually approaches the quark EoS above it, with the squared speed of sound exhibiting the same qualitative behavior. 

The location of the crossing point also has important consequences for the macroscopic properties of compact stars, with higher crossing chemical potentials being associated with larger maximum masses and radii.
Although other choices of $\mu_0$ and $r$ may produce mathematically well-defined crossover EoSs, they may fail to reproduce the underlying hadronic and quark phases and their crossing region reliably, potentially leading to physically misleading results. 
Finally, the relation obtained between the crossing chemical potential and the hadronic stiffness parameter $\eta$ provides a useful parametrization for predicting the crossing point for different hadronic EoSs. This relation can consequently be used to determine suitable values of $\mu_0$ and $r$ and thereby construct physically consistent crossover EoSs for a given hadronic parametrization.

An interesting prospect of the present approach is that, through an appropriate parametrization of the quark equation of state and the switching function, the main properties (mass and radius) of so-called exotic compact objects can be reproduced. These objects are characterized by relatively low masses and small radii, which lie outside the predictions of the standard model of neutron-star physics.

Future applications of the present study include using the derived equations of state to investigate the various oscillation modes (r-modes, f-modes, p-modes, g-modes etc.) that occur in the interior of compact objects. The most significant extension, however, involves the thermodynamic study of the equations of state by incorporating temperature into the calculations. This approach would enable us to investigate compact objects both at their formation, when they are still characterized by high temperatures, and during binary mergers, where the effects of the temperature generated in the merger process become non-negligible. A study along these lines is currently underway.







\end{document}